\documentclass[aps,pre,onecolumn,groupedaddress,longbibliography,nofootinbib,showkeys,showpacs,amssymb]{revtex4-1}

\usepackage{latexsym}
\usepackage{amsfonts}
\usepackage{graphicx}
\usepackage{mathptmx}      
\usepackage{bm}
\usepackage{enumerate}
\usepackage{subfigure}
\usepackage{color}
\usepackage{enumitem}

\newcommand\POS{SOC}

\DeclareRobustCommand\openone{\leavevmode\hbox{\small1\normalsize\kern-.33em1}}

\begin{document}


\title{Separation of conditions as a prerequisite for quantum theory}



\author{Hans De Raedt}
\affiliation{Zernike Institute for Advanced Materials,\\
University of Groningen, Nijenborgh 4, NL-9747AG Groningen, The Netherlands}
\author{Mikhail I. Katsnelson}
\affiliation{Radboud University, Institute for Molecules and Materials,
Heyendaalseweg 135, NL-6525AJ Nijmegen, The Netherlands}
\author{Dennis Willsch}
\affiliation{Institute for Advanced Simulation, J\"ulich Supercomputing Centre,\\
Forschungszentrum J\"ulich, D-52425 J\"ulich, Germany}
\author{Kristel Michielsen}
\thanks{Corresponding author}
\email{k.michielsen@fz-juelich.de}
\affiliation{Institute for Advanced Simulation, J\"ulich Supercomputing Centre,\\
Forschungszentrum J\"ulich, D-52425 J\"ulich, Germany}
\affiliation{RWTH Aachen University, D-52056 Aachen, Germany}

\date{\today}

\begin{abstract}
We introduce the notion of ``separation of conditions''
meaning that a description of statistical data obtained from experiments,
performed under a set of different conditions,
allows for a decomposition such that each partial description depends
on mutually exclusive subsets of these conditions.
Descriptions that allow a separation of conditions are shown to
entail the basic mathematical framework of quantum theory.
The Stern-Gerlach and the Einstein-Podolsky-Rosen-Bohm experiment with three, respectively nine
possible outcomes are used to illustrate how the separation of conditions
can be used to construct their quantum theoretical descriptions.
It is shown that the mathematical structure of separated descriptions implies that,
under certain restrictions, the time evolution of the data can be described by the von Neumann/Schr\"odinger equation.
\end{abstract}

\maketitle

\section{Introduction}\label{SPIN1}

Most of us heavily rely on our visual system to perform tasks in daily life.
The ability of our visual system to rapidly and effortlessly decompose a visual scene into
separate objects and categorize them according to their functionality
considerably enhances the chance of survival of the individual.
The example of the visual system is just one of the many instances
in which our cognitive system constantly performs separations ``on the fly''.
In daily life, we hardly notice that our brains are performing these separations,
suggesting that the basic processes involved are, as a result of evolution, hardwired into our brains.
Therefore, it is not a surprise that in many forms of cognitive activity,
also in the most abstract modes of human reasoning, separation into parts plays an important role.

There is a large variety problems in mathematics and physics for which separation
into parts is of great value.
For instance, separation of variables is a very powerful method for solving (partial) differential equations.
Describing the harmonic vibrations in solids in terms of normal modes (phonons) instead of
using the displacements of the atoms and their momenta is much more effective for
understanding their properties.
Analyzing a signal in terms of Fourier components is a standard method for decomposing
the signal into a sum of signals that each have a simple description.
Similarly, computing the principal components of a correlation matrix yields
a description of the data that, in many cases, is considerably simpler than the description of the data themselves.

The ubiquity of separation in cognitive processes suggests that
it may be an important guiding principle for developing useful descriptions of the phenomena that we observe.
In this paper, this guiding principle is used for the analysis and representation of data, as expressed by the statement
\begin{center}
\framebox{
\parbox[t]{0.9\hsize}{%
The separation of conditions (SOC),
when applied to data produced by experiments performed under several different sets of conditions
(e.g. $\{(a,b),(a',b')\}$),
reduces the complexity of describing the collective of these experiments
by decomposing the description of the whole into descriptions of several parts which
depend on mutually exclusive, proper subsets (e.g. $\{(a),(a')\}$ and $\{(b),(b')\}$) of the conditions only.
}}
\end{center}
It is important to recognize that \POS\ operates on a much more primitive level than e.g.
the principle of stationary action which is central in modern theoretical physics.
\POS\ serves as the foundation for a chain of reasoning
whereas the principle of stationary action refers to a general variational method that has numerous applications
across a wide field.
The latter principle is used to derive equations of motion from
a postulated functional called ``action''
whereas \POS\ is used by our cognitive system for a variety of functions.

It is remarkable that the evolution of our physical worldview goes hand in hand with
evolution of the main mathematical tools of theoretical physics.
Classical mechanics is based on the concept of materials points and enforces the use of ordinary
differential equations~\cite{NEWT99}.  According to Arnold~\cite{ARNO90}, the main achievement and
the main idea of Newton can be formulated in one sentence: ``It is useful to
solve (ordinary) differential equations''.
The Faraday-Maxwell revolution of 19th century placed the concept of field
in the center of theoretical physics, the corresponding
mathematical apparatus being partial differential equations~\cite{EINS67}.
Both the concepts of materials points and fields (e.g. water waves) relate directly to our daily experience~\cite{WEYL94}.
In contrast, in quantum theory, ``states'' of a system are vectors in a Hilbert space, ``observables'' are Hermitian
operators, and the mathematical apparatus is linear algebra and functional analysis~\cite{NEUM55}.
None of these concepts directly relates to elements of an experiment.
Numerous works on ``interpretation of quantum theory''
-- for a brief or concise overview of popular interpretations see Ref.~\onlinecite{WEIN03} or Ref.~\onlinecite{RAUC15},
respectively -- offer tens, hundreds of ways how to interpret the symbols of this language; much less
is known about its origin. Why is it that such abstract concepts play the central role
in our description of microscopic phenomena?
In this paper, we present an attempt to clarify this issue based on a
careful analysis of ways to organize information (represented by ``experimental data'')
and of ways to operate with it.

The view adopted in this paper is that the primary goal of a theoretical model should be to provide
concise descriptions of the available data
which constitute the objective information (i.e., free of personal judgment),
about the phenomena under scrutiny.
In addition, it is desirable to construct such descriptions using mathematics that is as simple as possible.
Due to the general, non-mathematical nature of \POS, it is impossible to deduce or derive, in the mathematical sense,
the basic postulates of quantum theory from \POS\ only: one has to inject into the mathematical framework that
is constructed on the basis of \POS, additional knowledge about the specific conditions under which
the experiments are being performed.
In this paper, we adopt the traditional approach of theoretical physics by assuming
that the phenomena under scrutiny allow for a continuum space-time description.
In other words, the additional assumption that we will use (implicitly) is that, in mathematical terms,
the symmetries of the space-time continuum apply.
Furthermore, as discussed later in this paper, quantum theory is not the only theory
consistent with \POS, a statement which we will symbolically denote as
\begin{equation}
\mathrm{SOC}\models \mathrm{QT}
,
\label{QTSOC}
\end{equation}
meaning that models in $\mathrm{SOC}$ entail models of $\mathrm{QT}$.
However, the application of \POS\ yields a framework that is specific enough
such that the language of operators and state vectors appears in a natural manner, consistent with our experience
that experiments count individual events.
Phrased more concisely, we show that quantum theory is a model for the specific class of data (of frequencies of events)
to which \POS\ applies.

\subsection{Quantum physics experiments}

Consider a typical scattering experiment in which a crystal is targeted
by neutrons produced by a nuclear reactor.
Obviously, a description of the experiment as a whole is much too complicated to be useful.
Therefore, we simplify matters. First, we leave out the description of the whole nuclear reactor as a neutron source
and imagine a fictitious source preparing neutrons with well-defined momenta.
Next, we assume that we know how to model the interaction of the neutrons with the atoms of the crystals.
The measurement itself consists of detecting, one-by-one, the neutrons that leave the crystal in various directions.
Finally, interpreting the counts of the neutrons scattered by the crystal in terms of the
neutron-crystal interaction model allows us to make inferences about the lattice or magnetic structure of the crystal.

From this rough sketch of the neutron scattering experiment, it is clear that
\POS\ has been used before any attempt is made to describe the experiment by a mathematical model.
\POS\ seems to be an (implicit) assumption of all physical theories that have been invented.
In particular, standard introductions of the quantum formalism
assume -- and do so often implicitly --  that a model of the experiment
can be decomposed into a preparation stage and a measurement stage,
before the first postulate is introduced, see e.g. Ref.~\cite{BALL03}.

A characteristic feature of quantum physics experiments such as the neutron scattering experiment sketched earlier
is the uncertainty in behavior of the individual neutrons.
In essence, single events are regarded as irreproducible.
Moreover, because the observed counts are the basis for making inferences about the interaction model,
these inferences may be subject to additional uncertainties, in addition to those due to the uncertainties on the incoming neutrons.
If single events are not reproducible, we may have (but not necessarily have) the situation that
in the long run, the relative frequencies of the different detection events approach reproducible numbers.
The latter means that upon repetition of the whole experiment, i.e. by collecting the data of many detection events,
deviations of the new relative frequencies from the previous ones are within the statistical errors,
e.g. they satisfy the law of large numbers~\cite{BARL89,GRIM01}.

Results of laboratory experiments are always subject to uncertainties.
In the theoretical description of these results we may choose to ignore these uncertainties,
for good reason as in Newtonian mechanics, or not, as in quantum theory.
The latter has no means by which to calculate the outcome of an individual event, a feature
it shares with Kolmogorov's probability theory~\cite{KOLM56,GRIM01}.
Not being able to deduce from a theory the very existence
of the individual events that we observe is at the heart of the difficulties of understanding
what the theory is about and what it describes.

\subsection{Quantum theory}

Quantum theory is probably the most obscure and impenetrable subject of all current scientific theories.
Text books on quantum theory usually start with a brief historical account of the experiments
that were crucial for the development of the theory to make plausible the postulates which
define the mathematical framework~\cite{BOHM51,FEYN65,BALL70,KHRE09,BALL03}.
Subsequent efforts then go into mastering linear algebra in Hilbert space, solving partial differential equations, and
other abstract mathematical tools.
The tendency to focus on the elegant mathematical formalism~\cite{NEUM55},
which, unfortunately, is far more detached from everyday experience than for instance Newtonian mechanics or electrodynamics,
promotes the ``shut-up-and-calculate'' approach~\cite{RALS17}.
Hermitian operators, wave functions, and Hilbert spaces are conceptual, mental constructs which have no tangible counterpart
in the world as we experience it through our senses.
The mathematical results that are derived from the postulates of a theoretical model are only theorems within the axiomatic
framework of that theoretical model.
Theoretical physics uses axiomatic frameworks which have a rich mathematical structure, allowing the proof of theorems.
For instance, the Banach-Tarsky paradox~\cite{BANA24} has no counterpart in the world that humans experience.
Taking the mathematical description for real is like opening bottles that contain very exotic and sometimes magical substances.
In other words, relating theorems derived within a mathematical axiomatic formalism to observable reality is not a trivial matter.

In quantum theory, the mapping from what occurs in Hilbert space to what is taking place
in the laboratory is further convoluted by the fact that quantum theory lacks the means to account for the fact that a single measurement
has a definite outcome~\cite{BOHM51,FEYN65,BALL03}.
That is, quantum theory cannot describe the fact that humans register individual events
although it does a wonderful job to describe, under appropriate conditions, the frequencies with which these events occur.
The conundrum of not being able to deduce from the theory that each measurement yields a definite outcome~\cite{ALLA17}
manifests itself in the number of different quantum-theory interpretations that exist today.

It may be of interest to mention here that the formalism of quantum theory finds
applications in fields of science that are not even remotely related to the physics experiments which
cannot be described by classical physics~\cite{KHRE10}.
This begs the question ``Why is the quantum formalism also useful in these non-quantum applications?''.

\subsection{Application of \POS}

In this paper, we explore a route, based on \POS, to construct the quantum theoretical description
without running into the conundrum mentioned earlier.
We start from the empirical fact that the result of a measurement yields a definite outcome.
We review several different, simple ways to represent the moments of the relative frequencies of the different outcomes.
It then follows that the mathematical structure underlying quantum theory is the simplest of
many equivalent representations that allows the description of the experiment
to be decomposed into a description of a preparation stage and a measurement stage
(an  implicit assumption in the formulation of quantum theory~\cite{BALL03}).

It is obvious that this way of thinking is opposite to the more traditional, deductive reasoning
which assumes an underlying ontology~\cite{BOHM52,PENA96,THOO97,PENA05,THOO07}
or starts from various, different sets of axioms~\cite{%
LAND74,FRIE89,VSTO95,REGI98,HARD01,LUO02,FRIE04,BUB07,CAVE07,PALG08,KHRE09,KAPS10,CHIR11,MASA11,BRUK11,%
SKAL11,KAPS11,ORES12,SANT12,KLEI12a,FLEG12,KAPU13,FUCH13,HOLI14},
the individual event being the last (but apparently unreachable) element in the chain of thoughts.
This is also evident from the fact that in our construction there is no need to
even mention the concept of probability, simply because the mathematical structure
directly follows from a rearrangement of the data (counts of events) and the application of \POS.
For a different approach based on rearranging data, see Ref.~\onlinecite{RALS13b}.

Another approach which reverses the chain of thought, i.e. starts from the notion of an individual event,
uses the algebra of logical inference (LI), a mathematical framework for
rational reasoning in the presence of uncertainty~\cite{COX46,COX61,TRIB69,SMIT89,JAYN03}.
Applying LI to reproducible and robust experiments yields a description in terms of a seemingly complicated
nonlinear global optimization problem, the solutions of which can be shown to be equivalent
to the extrema of a quadratic form.
For instance, for one particular scenario of collecting data, we recover the (time-dependent)
Schr\"odinger equation~\cite{RAED14b,RAED16b}.
Similary, for two other scenarios, LI yields the Pauli or Klein-Gordon equations, respectively,
all without invoking concepts quantum theory~\cite{RAED15b,DONK16}.

The \POS\ approach only yields the basic mathematical framework, in essence only the postulates that appear in the
statistical (ensemble) interpretation of quantum theory~\cite{BALL70,BALL03} (see below), knowledge about the specific physical
problem has to be supplied in terms of symmetries, the correspondence principle etc., as in conventional quantum theory.
In contrast, the LI approach starts from the notion of reproducible and robust frequencies of events, uses
the requirement that in the absence of uncertainty classical Hamiltonian mechanics is recovered,
and allows us to derive, in a strict mathematical sense, e.g. the Schr\"odinger equation of the hydrogen atom.
So far, all equations derived on the basis of LI describe quantum systems in a pure state only~\cite{RAED13b,RAED14b,RAED15b,RAED15c,RAED16b,DONK16}.
Whether the LI approach can be extended to cover quantum systems in a mixed state is an open problem.

Recently, we have given simple examples, one of them in the context of the EPRB experiment,
showing that LI can describe realizable data sets which do not
allow for a quantum theoretical description~\cite{RAED18a}.
Therefore, symbolically we have
\begin{equation}
\mathrm{LI} \models \mathrm{QT}^\ast
.
\label{QTLI}
\end{equation}
where $\mathrm{QT}^\ast$ denotes the
class of quantum systems characterized by pure states.
On the other hand, the \POS\ approach contains the results of the LI approach, such that
\begin{equation}
\mathrm{SOC} \models \mathrm{LI} \models \mathrm{QT}^\ast
\quad\mathrm{and}\quad
\mathrm{SOC} \models \mathrm{QT}
.
\label{QTLISOC}
\end{equation}

\begin{figure}[t]
\centering
\includegraphics[width=0.8\hsize]{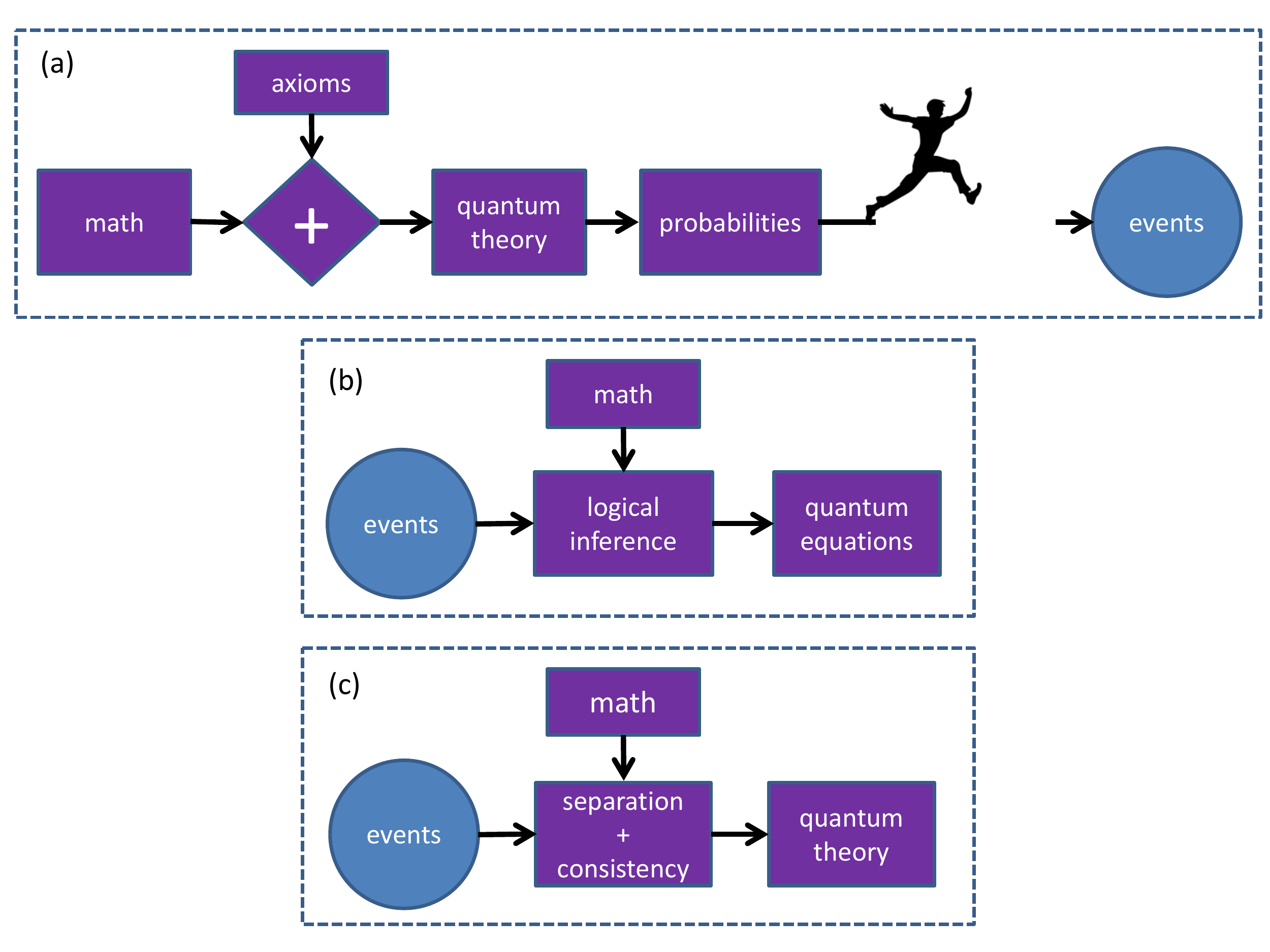}  
\caption{(Color online)
Diagram illustrating the different chains of thought.
(a) Traditional method to formulate quantum theory;
(b) logical inference approach for deriving basic equations of quantum theory;
(c) the approach explored in this paper.
In contrast to the traditional method, which lacks the means to predict the
individual, observable event, the other two approaches put the event at the
core of reasoning.
In (a) and (c) the box ``quantum theory'' refers to the mathematical framework,
defined by postulates P1 and P2.
}
\label{nopostulates}
\end{figure}

As already mentioned, this paper takes the notion of an individual event as the starting point
but does not require the experiment to be reproducible nor to be robust.
The only requirement is that the data gathered in one run
is completely described by the relative frequencies of the different kinds of detection events,
a number of which, in any laboratory experiment, is always finite.
We show that \POS\ and consistency, combined with a simple reorganization
of the observed data and some standard assumptions (e.g. symmetries of the space-time continuum ),
are sufficient to construct the mathematical framework of quantum theory for a finite number of different outcomes.
By consistency we mean that the description
of a particular part is independent of the experiment or context in which the part is used.

A graphical representation of the traditional, the logical inference,
and the \POS\ approach
to introduce the formalism of quantum-theory is shown in Fig.~\ref{nopostulates}.
As far as we know, all interpretations of quantum theory are based on the same
expressions for expectation values of dynamical variables such as position, energy etc.
The difference between interpretations appears in the way the theoretical description deals with the measurement problem,
i.e. ``explains'' that each measurement yields a definite outcome.
The statistical (ensemble) interpretation of quantum theory is silent about this aspect.
Copenhagen-like interpretations postulate the elusive
wave function collapse to ``explain'' the existence of events.

Independent of the interpretation that one prefers,
there is the crucial fact, almost never mentioned, that
a genuine probabilistic theory does not entail a procedure or process by which elementary events can actually be produced.
The existence of a set of elementary events is assumed, and probability theory is then built on this assumption~\cite{KOLM56}.
Ways to produce events according to a specified probability distribution would be
(1) call Tyche to produce events without undiscoverable cause, i.e. appeal to magic,
or (2) use an algorithm to let a computer generate events.
Obviously, the latter is deterministic, pseudo-random in nature, does not produce random events
in the strict mathematical sense, and  is ``outside'' probability theory.

The two other approaches, graphically represented in Fig.~\ref{nopostulates}(b,c), do not
suffer from the problem of not being able to generate events.
Indeed, in both the logical inference and the \POS\ approach,
the event is the key element on which the whole theoretical structure is built.
There is no need to have a procedure to generate events
according to a specified probability distribution.
Instead, this distribution is constructed from the frequencies of the events
(and additional pieces of knowledge, depending on the case at hand).

Instead of discussing the application of \POS\ to quantum physics experiments in its most general form,
we choose the more instructive route by demonstrating its application to two simple,
but non-trivial experiments which have been instrumental in the development of quantum theory.
The mathematical framework that emerges from applying \POS\ generalizes in an almost trivial manner.
Following Feynman~\cite{FEYN65}, we use the Stern-Gerlach (SG) experiment to illustrate how
its quantum theoretical description directly emerges from a representation of the observed data in terms
of independent, separate descriptions of the source and the SG magnet.
We explicitly show that \POS\ in combination with
the requirement of consistency
and the use of symmetries of the space-time continuum suffice to recover
the quantum theoretical description of a spin one ($S=1$) system.
As a further illustration, we consider the Einstein-Podolsky-Rosen-Bohm experiment (EPRB)
and show how also in this case the quantum theoretical description derives from
a representation of the observed data in terms of independent, separate descriptions of the source and SG magnets.
This example also demonstrates how to extend the approach to many-body problems.
The work presented in this paper extends and generalizes our earlier work~\cite{RAED15c,RAED16b} on the spin-1/2 case.

\subsection{Preview of the main result}

In general terms, the main result of this paper can be summarized as follows.
The mathematical structure of the following two postulates (or equivalent formulations of them)

\medskip\noindent
P1. {\sl To each dynamical variable R (physical concept) there corresponds
a linear operator $R$ (mathematical object), and the possible values of the
dynamical variable are the eigenvalues of the operator}~\cite{BALL03}.

\medskip\noindent
and

\medskip\noindent
P2. {\sl To each state there corresponds a unique state operator. The
average value of a dynamical variable R, represented by the operator $R$, in
the virtual ensemble of events that may result from a preparation procedure
for the state, represented by the operator $\rho$, is
$\langle R\rangle = \mathbf{Tr\;}\rho R /\mathbf{Tr\;}\rho$}~\cite{BALL03}.

\medskip\noindent
which form the basis for the statistical (ensemble) interpretation of quantum theory~\cite{BALL70,BALL03}
and suffice for all practical ``shut-up-and-calculate'' applications of quantum theory,
directly follow from the application of \POS\ and a simple rearrangement of the data for the frequencies of the observed events.
Note that neither quantum theory nor \POS\ yield the expressions of $\rho$ or $R$.
Obviously, these expressions depend on the details of the experiment.
Application of \POS\ to data gathered in quantum physics experiments
provides an answer to the riddle ``Where does the quantum formalism come from and why is it useful
in non-quantum applications?''.

\subsection{Structure of the paper}
The paper is organized as follows.
In Sec.~\ref{sec2}, we sketch the experimental setup of the double SG
experiment that we use as the primary example to illustrate the
application of \POS\ to data obtained by performing experiments under different conditions.
Section~\ref{sec2a} discusses the kind of data that are generated by this experiment and
their characterization in terms of moments.
In Sec.~\ref{sec2e}, we introduce \POS\ using the SG experiment
with three different outcomes as an example and show that matrix algebra
allows for the description to be separated in the sense of \POS.
Explicit expressions for the description of the measurement stage are given in Secs.~\ref{sec2f} and \ref{sec2g}.
The application to the double SG experiment, given in Sec.~\ref{sec2h}, completes the construction
and also shows how the basic structure of the quantum formalism emerges from \POS.
In Sec.~\ref{sec2i}, we work out in detail a specific example of the double SG experiment and
show that quantum theory restricts the functional dependence of the observed frequencies on the SG magnet parameters
to those dependences for which separation is possible.
Section~\ref{sec2j} discusses the most general description of the particle source and
also the measurements that are required to fully characterize this source.
Application of \POS\ enforces a representation of the data in terms of matrices,
suggesting that there may be a relation to Heisenberg's matrix mechanics~\cite{HEIS25}.
In Sec.~\ref{sec9}, we scrutinize this relation and argue that if there is one, it is very weak.
Section~\ref{sec4} explores the conditions under which
the time evolution of data that allows for a separated description
can be described by the von Neumann/Schr\"odinger equation.
Using the EPRB experiment as the simplest, nontrivial example, we demonstrate in Sec.~\ref{sec3}
how the tensor-product structure of quantum many-body physics naturally
emerges from the application of \POS.
In Sec.~\ref{sec11}, we discuss the general features of the \POS\ construction
of the quantum formalism and its relation to the commonly accepted postulates of quantum theory.
Our conclusions are given in Sec.~\ref{sec10}.

\section{Double Stern-Gerlach experiment}\label{sec2}

The SG experiment~\cite{STER22,GERL24,BOHM51,FEYN65}
involves sending particles through an inhomogeneous magnetic field and observing their deflection.
A source emits particles such as
atoms~\cite{STER22,GERL24}, neutrons~\cite{SHER54,HAME75}, electrons~\cite{BATE97}, or atomic clusters~\cite{BACH18}.
Particles are sent one-by-one through a SG magnet, the salient feature of which is
that it generates an inhomogeneous magnetic field,
along a direction characterized by the unit vector $\mathbf{a}$.
The interaction of this field with magnetic moment of the particles changes the momentum of the latter.
As a result, the particle beam is split into
in $2S+1$ spatially well-separated directions which are determined by the unit vector $\mathbf{a}$,
an experimental fact~\cite{STER22,SHER54,HAME75,BACH18}.
This experimental fact is regarded as direct evidence for the quantized magnetic moment~\cite{STER22,BOHM51,FEYN65}.
The latter is proportional to the ``spin'' of the particle and is assigned a magnitude $S$.

Assume that it is already established by experiments that there is a magnetic but no electric field
between the poles of a SG magnet and that it is known, also from experiments, that
the particles under scrutiny do not carry electrical charge.
Then, if these particles pass through the SG magnet and show a deflection
that is absent when the magnetic field is zero, it makes sense to assign the attribute ``magnetic''
to these particles.
The observed deflection can be attributed to the interaction between
the magnetic field inside the SG magnet and the assigned magnetic quality of the particles.

We now wish to go a step further and assign to the particles a definite magnetic moment, characterized by a
direction and size, a necessary step if we want to speak about quantized magnetic moments.

\begin{figure}[h]
\centering
\includegraphics[width=0.8\hsize]{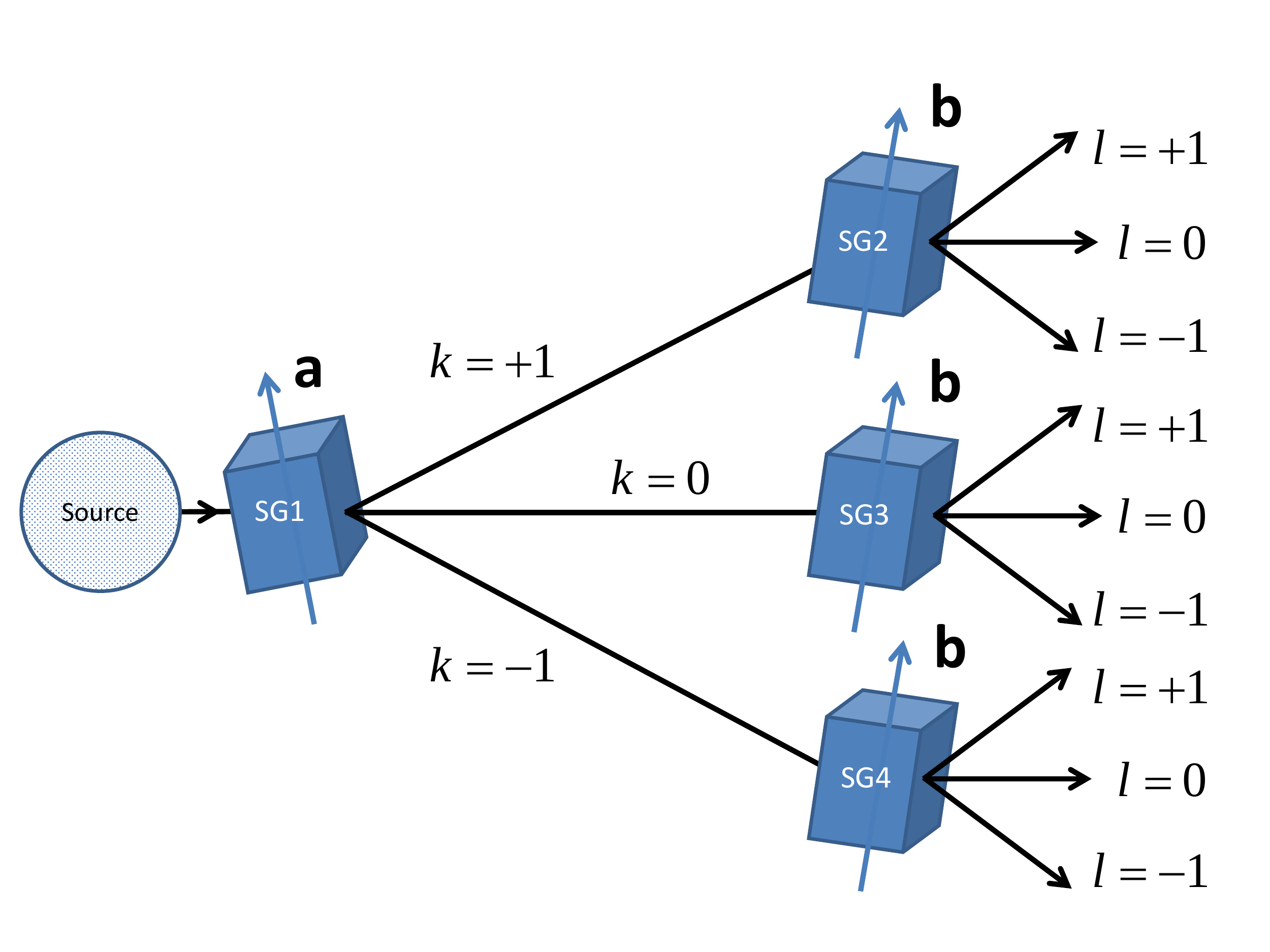}  
\caption{(Color online)
Layout of the double Stern-Gerlach experiment that we consider in this paper.
Electrically neutral, magnetic particles leave the source one-by-one
and pass through the inhomogeneous magnetic field,
characterized by the same unit vector $\mathbf{a}$,
created by the Stern-Gerlach magnet SG1.
Particles leave SG1 in one of the three beams labeled by $k=+1,0,-1$.
The direction of these beams depends on the unit vector $\mathbf{a}$.
Particles then travel to either SG2, SG3, or SG4, all three of them
characterized by the same unit vector $\mathbf{b}$.
Particles leave SG2, SG3, or SG4 in one of the three beams labeled by $l=+1,0,-1$.
Finally, each particle is registered by one and only one detector (detectors not shown)
labeled by $(k,l)$.
}
\label{fig0}
\end{figure}

In general, a consistent assignment of a particle property is a two-step process.
First, we employ a filter to select particles.
Then, using a second {\bf identical} filter, we verify that all the particles pass that second filter.
In the case at hand, this procedure amounts to performing a double SG experiment~\cite{BOHM51,FEYN65}
such as the one sketched in Fig.~\ref{fig0}.
Only if the direction of the magnetic moment is ``preserved'' during repeated probing, it makes
sense to attribute to the particle, a definite direction of magnetization.

We do not know of any laboratory realization of a double SG experiment but,
following Feynman~\cite{FEYN65}, we use this thought experiment to construct the theoretical description
which, in contrast to Feynman's approach, does not build on postulates of quantum theory.
Also following Feynman~\cite{FEYN65}, we focus on the case
of particles which, in quantum parlance, are said to have spin $S=1$.
Generalization to other values of the spin is straightforward~\cite{FEYN65}.
We emphasize that our choice to illustrate the main ideas by using experiments
with three outcomes per SG magnet is only for the sake of balance between generality and simplicity.
Our treatment readily generalizes to experiments with any number of different outcomes.

Following standard practice in developing theoretical models, we assume that the experiment is ``perfect'' in
the sense that all SG magnets are identical, their inhomogeneous magnetic fields are constant for the duration
of the experiment, each particle leaving the source is detected by one and only one
of the nine detectors, and so forth.

\section{Data generated by the experiment}{\label{sec2a}}

As is clear from Fig.~\ref{fig0}, for each particle leaving the source,
one and only one detector, labeled by $(k,l)$, will fire.
It may be tempting to say that the detected particle traveled
along the beams labeled by $k$ and $l$ but as a matter of fact,
on the basis of the available data, i.e. $(k,l)$, no such assignment can be made.
On the other hand, for the purposes of this paper, it does no harm to {\sl imagine}
and it also simplifies the writing to say that the particle
followed a particular path, but to know this for sure, we would have
to add detectors in the beams between the SG1 and second layer of SG magnet.

We start by considering the experiment in which the second layer (SG2, SG3, SG4)
is absent and detectors are placed in the beams labeled by $k$.
We denote the counts of detector clicks recorded after $N$ particles have left the source by
\begin{equation}
{\cal K}=\big\{ k_n\;|\; k_n\in\{+1,0,-1\}\;;\; n=1,\ldots ,N\big\}
.
\label{sec2a1}
\end{equation}

The relative frequency with which particles travel along the path $k$ is given by
\begin{eqnarray}
f(k|\mathbf{a},P,N)=\frac{1}{N}\sum_{n=1}^N \delta_{k,k_n}
.
\label{sec2a2b}
\end{eqnarray}
We introduce the notation $|\mathbf{a},P,N)$ to indicate that the $N$ data items have been collected
during a period in which $\mathbf{a}$ and the properties of the particles,
represented by the symbol $P$, are assumed to be constant.

Similarly, for the double SG experiment, repeating the experiment with $N$ particles yields the data set
\begin{equation}
{\cal D}=\big\{ (k_n,l_n)\;|\; k_n,l_n\in\{+1,0,-1\}\;;\; n=1,\ldots ,N\big\}
,
\label{sec2a0}
\end{equation}
and the relative frequency with which particles travel along the paths $(k,l)$ is given by
\begin{eqnarray}
f(k,l|\mathbf{a},\mathbf{b},P,N) =\frac{1}{N}\sum_{n=1}^N \delta_{k,k_n}\delta_{l,l_n}
.
\label{sec2a2a}
\end{eqnarray}
In Eq.~(\ref{sec2a2a}), $|\mathbf{a},\mathbf{b},P,N)$ indicates that the $N$ data items have been collected
during a period in which $\mathbf{a}$ and $\mathbf{b}$ and the properties of the particles,
represented by the symbol $P$ are assumed to be constant.
Regarding the meaning of $P$, it is important to note that properties of the particles under scrutiny
can only be assigned a-posteriori on the basis of experimental data.

Obviously, relative frequencies do not contain information about correlations between events, if any were present.
Therefore, in general, a full characterization of data in the set ${\cal D}$ (${\cal K}$)
requires more than just the knowledge of the relative frequencies $f(k,l|\mathbf{a},\mathbf{b},P,N)$ ($f(k|\mathbf{a},P,N)$).
\begin{center}
\framebox{
\parbox[t]{0.8\hsize}{%
In this paper, we only analyze the simplest case by
discarding all knowledge about the events that is not contained in
the relative frequencies.
}}
\end{center}

For later use, we write $f(k|\mathbf{a},P,N)$ in terms of its moments defined by
\begin{eqnarray}
m_p(\mathbf{a},P,N)=\langle k^p \rangle_{\mathbf{a}}=\frac{1}{N}\sum_{n=1}^N k_n^p = \sum_{k=+1,0,-1} k^p f(k|\mathbf{a},P,N)
\quad,\quad p=0,1,2,
\label{sec2a2c}
\end{eqnarray}
where, by construction, the zero'th moment is $m_0=1$ and
$\langle X \rangle_{\mathbf{a}}$ denotes the average of $X$ with respect to the relative frequencies $f(k|\mathbf{a},P,N)$.
The explicit expression of $f(k|\mathbf{a},P,N)$ in terms of its moments $m_0$, $m_1$ and $m_2$
can be found by solving the corresponding linear set of equations.
We have
\begin{eqnarray}
f(k|\mathbf{a},P,N) &=& 1-m_2(\mathbf{a},P,N) + \frac{m_1(\mathbf{a},P,N)}{2}  k + \frac{3m_2(\mathbf{a},P,N)-2}{2} k^2
,
\label{sec2a3}
\end{eqnarray}
which is consistent with  Eq.~(\ref{sec2a2c}).

According to the boxed text above, in this paper we
take the viewpoint that data in ${\cal K}$
is completely described by two relative frequencies, e.g.
$f(-1|\mathbf{a},P,N)$ and $f(0|\mathbf{a},P,N)$, and the normalization
$f(-1|\mathbf{a},P,N)+f(0|\mathbf{a},P,N)+f(+1|\mathbf{a},P,N)=1$
or, equivalently, by the moments
$m_0=1$, $m_1(\mathbf{a},P,N)$, $m_2(\mathbf{a},P,N)$, and Eq.~(\ref{sec2a3}).
Similarly, the data in ${\cal D}$ is completely described by the
moments $\langle k^p l^q\rangle_{\mathbf{a}}$ for $p,q=0,1,2$.

\section{Application of \POS}{\label{sec2e}}

Given the description of the data ${\cal K}$ in terms of relative frequencies
$f(k|\mathbf{a},P,N)$, we ask ourselves whether it
is possible to apply the general idea of separation to the SG experiment
and construct a description of the whole in terms of descriptions
of the various components of the experiment, in the case at hand the particle source and the SG magnet.
That such a separation can be made was already shown for the spin-1/2 SG and Bell-type experiments~\cite{RAED15c,RAED16b}.
In this paper, we show that this approach extends to higher spin and leads to the same
conclusion~\cite{RAED15c,RAED16b}, namely that a separation is possible if we write the same data
in matrix rather than in vector form.

We begin by focusing on the first stage of the double SG experiment depicted in Fig.~\ref{fig0} and
do some innocent-looking rewriting.
Let us organize the observations $(+1,0,-1)$ and the relative frequencies into vectors $\mathbf{k}=(+1,0,-1)^T$
and $\mathbf{f}=(f(+1|\mathbf{a},P,N),f(0|\mathbf{a},P,N),f(-1|\mathbf{a},P,N))^T$, respectively.
We have
\begin{eqnarray}
\langle 1 \rangle_{\mathbf{a}} &=& (1,1,1)\cdot\mathbf{f}
=\mathbf{Tr\;}(1,1,1)\cdot\mathbf{f}
=\mathbf{Tr\;}\mathbf{f}\cdot (1,1,1)
=\mathbf{Tr\;}
\left(\begin{array}{ccc}
f(+1|\mathbf{a},P,N) & 0 & 0\\
0& f(0|\mathbf{a},P,N)&0\\
0&0&f(-1|\mathbf{a},P,N)\\
\end{array}\right)
,
\label{sec2e0a}
\\
\noalign{\noindent and}
\langle k \rangle_{\mathbf{a}} &=& \mathbf{k}^T\cdot\mathbf{f}
=\mathbf{Tr\;}\mathbf{k}^T\cdot\mathbf{f}
=\mathbf{Tr\;}\mathbf{f}\cdot \mathbf{k}^T
=\mathbf{Tr\;}
\left(\begin{array}{ccc}
f(+1|\mathbf{a},P,N) & 0 & 0\\
0& 0&0\\
0&0&-f(-1|\mathbf{a},P,N)\\
\end{array}\right)
,
\label{sec2e0b}
\end{eqnarray}
where $\mathbf{Tr}\; \mathbf{A} $ denotes the trace of the matrix $\mathbf{A}$, i.e. the sum of all diagonal elements of $\mathbf{A}$,
and we made use of the invariance of the trace under cyclic permutation of the matrices, i.e.
$\mathbf{Tr}\; \mathbf{A}\mathbf{B} = \mathbf{Tr}\; \mathbf{B}\mathbf{A}$.
Equations~(\ref{sec2e0a}) and (\ref{sec2e0b}) express the normalization condition and the average of $k$
as the trace of the $3\times3$ matrices $\mathbf{f}\cdot (1,1,1)$ and $\mathbf{f}\cdot\mathbf{k}^T$.

It is not possible to write down an expression similar to Eq.~(\ref{sec2e0b}) that yields
$\langle k^2 \rangle_{\mathbf{a}}$ unless we introduce a new vector $\mathbf{k}^{(2)}=(+1,0,+1)^T$
and define
$\langle k^2 \rangle_{\mathbf{a}} = \mathbf{Tr\;}\mathbf{f}\cdot(\mathbf{k}^{(2)})^T$.
However, if we write the observations and relative frequencies as $3\times3$ diagonal matrices
\begin{eqnarray}
\mathbf{\widetilde K}=
\left(\begin{array}{rrr}
+1&\phantom{+}0&\phantom{+}0\\
\phantom{+}0&0&0\\
0&0&-1\\
\end{array}\right)
\quad\mathrm{and}\quad
\mathbf{\widetilde F}({\mathbf{a}},P,N)=\left(\begin{array}{ccc}
f(+1|\mathbf{a},P,N)&0&0\\
0&f(0|\mathbf{a},P,N)&0\\
0&0&f(-1|\mathbf{a},P,N)\\
\end{array}\right)
,
\label{sec2e1}
\end{eqnarray}
respectively, we have as a result of standard matrix algebra that
\begin{eqnarray}
\langle k^p \rangle_{\mathbf{a}} = \mathbf{Tr\;}\mathbf{\widetilde F}({\mathbf{a}},P,N)\mathbf{\widetilde K}^p
\quad,\quad p=0,1,2
.
\label{sec2e2}
\end{eqnarray}
Thus, using representation Eq.~(\ref{sec2e1}), there is no need to introduce an object (such as $\mathbf{k}^{(2)}$) to
represent $\langle k^2 \rangle$.
Note that a similar argument played a key role in Heisenberg's construction of his matrix mechanics~\cite{HEIS25}.

Up to this point, rewriting Eq.~(\ref{sec2a2c}) as Eqs.~(\ref{sec2e1}) and (\ref{sec2e2}) does not seem to bring anything new.
However, as we now show, by arranging numbers in matrices instead of vectors,
it becomes possible to perform the desired separation in
terms of a description of the source and the SG magnet~\cite{RAED15c,RAED16b}.
The key idea is to note that {\bf any} pair of matrices $\mathbf{F}$ and $\mathbf{K}$ satisfying
\begin{eqnarray}
\langle k^p \rangle_{\mathbf{a}} = \mathbf{Tr\;}\mathbf{F}\mathbf{K}^p
\quad,\quad p=0,1,2,
\label{sec2e3}
\end{eqnarray}
is a valid and therefore potentially useful representation of the data set ${\cal K}$, see Eq.~(\ref{sec2a3}).
As will become clear later on, it is not a coincidence that Eq.~(\ref{sec2e3}) resembles the expression
of an expectation value of a system in a quantum state described by a density matrix.

From Eq.~(\ref{sec2e3}) it is clear that the only way to separate
the description of the source from that of the SG magnet is to require that
the former, i.e. $\mathbf{F}$, does not depend on the
direction of the magnetic field ${\mathbf{a}}$
whereas the latter, i.e. $\mathbf{K}$, does.
We make this explicit by writing $\mathbf{K}({\mathbf{a}})$ in the following
and rewrite Eq.~(\ref{sec2e3}) as
\begin{eqnarray}
\langle k^p \rangle = \mathbf{Tr\;}\mathbf{F}(P,N)\mathbf{K}^p({\mathbf{a}})
\quad,\quad p=0,1,2,
\label{sec2e3a}
\end{eqnarray}
where we dropped the subscript in $\langle . \rangle_{\mathbf{a}}$
to emphasize that $\langle . \rangle$ refers to averages with respect to the matrix $\mathbf{F}(P,N)$
which does not depend on ${\mathbf{a}}$.

The left-hand side of Eq.~(\ref{sec2e3a}) is obtained by counting events and is, for each $p$, a rational number.
Therefore, we should impose that $\mathbf{Tr\;}\mathbf{F}(P,N)\mathbf{K}^p({\mathbf{a}})$ is real-valued,
but there is no such constraint on the matrices $\mathbf{F}(P,N)$ or $\mathbf{K}^p({\mathbf{a}})$.
For $p=0$, this implies that  $\mathbf{Tr\;}\mathbf{F}(P,N)=\mathbf{Tr\;}\mathbf{F}^\dagger(P,N)$
where, as usual, ``$^\dagger$'' stands for Hermitian conjugate.
This requirement is satisfied if $\mathbf{F}(P,N)$ is Hermitian
but $\mathbf{F}^\dagger(P,N)=\mathbf{F}(P,N) + \mathbf{X}$
with $\mathbf{Tr\;}\mathbf{X}=0$ would be allowed too.

An obvious route to search for the pair $(\mathbf{F}(P,N),\mathbf{K}({\mathbf{a}}))$ is to
use the property that the trace of a matrix does not change under a similarity transformation $\mathbf{R}$.
Thus, looking for matrices $\mathbf{R}$ such that
$\mathbf{F}(P,N)=\mathbf{R}\mathbf{\widetilde F}({\mathbf{R}})\mathbf{R}^{-1}$
and
$\mathbf{K}({\mathbf{R}})=\mathbf{R}\mathbf{\widetilde K}\mathbf{R}^{-1}$
might seem a viable route to explore.
However, limiting the search to similarity transformations is overly restrictive because
it does not allow for transformations of the kind
$\mathbf{\widetilde F}({\mathbf{R}})\mathbf{\widetilde K}^p =
\mathbf{F}(P,N)\mathbf{K}^p({\mathbf{R}}) + \mathbf{X}$
where $\mathbf{X}$ is a matrix of trace zero.
In fact, for the spin-1/2 case, the transformation that produces the desired separation is of this type~\cite{RAED15c,RAED16b}.
In summary, the requirement that only the traces of the matrices
should not change if we switch from representation Eq.~(\ref{sec2a3})
to Eq.~(\ref{sec2e3a}) still leaves a lot of freedom in the choice of the representation.

We would like to emphasize that
\begin{center}
\framebox{
\parbox[t]{0.9\hsize}{%
\begin{enumerate}
\item
Equations~(\ref{sec2e1})--(\ref{sec2e3a}) are not postulated but are instead obtained by a simple rewriting of
two sets of numbers as two square arrays instead of two linear lists
and by noting that there is considerable flexibility in choosing the arrays.
\item
There is, a-priori, no reason why $\langle k^p \rangle_{\mathbf{a}}$ for $p=1,2$
allows for a separation of the form Eq.~(\ref{sec2e3a}).
\item
In this particular example, \POS\ splits the compound condition
$(\mathbf{a},P,N)$ into the conditions $(\mathbf{a})$ and $(P,N)$.
\item
If \POS\ applies, the data gathered in the SG experiment
(i.e. not the imagined data represented in terms of real numbers) can be expressed in the form
Eq.~(\ref{sec2e3a}) which has the mathematical structure of postulate P2 of quantum theory.
\item
Up to this point in the paper, all variables take rational values only.
Starting from Eq.~(\ref{sec2e3a}) one cannot derive, in a strict mathematical sense, a theoretical framework
that uses irrational, real, or complex numbers but, as is well-known from number theory,
one can construct such a framework by an appropriate limiting process.
In the sections that follow, we bypass such a construction
by adopting the traditional viewpoint of theoretical physics that space-time is a continuum
and use complex numbers for convenience.
\end{enumerate}
}}
\end{center}

\section{Explicit form of $\mathbf{K}({\mathbf{a}})$}{\label{sec2f}}

Suppose that initially, the particles travel in the $x$-direction
and that ${\mathbf{a}}$ is along the $z$-direction, both directions being
fixed with respect to the laboratory frame of reference $(\mathbf{e}_x,\mathbf{e}_y,\mathbf{e}_z)$.
Then, the deflection of a particle that ends up in the $k=+1$ and $k=-1$ beam
can be associated with the $+{\mathbf{e}}_z$ and $-{\mathbf{e}}_z$ direction, respectively.
In other words, $\mathbf{K}({\mathbf{e}}_z)$ is just the matrix $\mathbf{\widetilde K}$
given in Eq.~(\ref{sec2e1}).
The expression of $\mathbf{K}({\mathbf{a}})$ is then readily found by
performing the rotation that turns $\mathbf{e}_z$ into $\mathbf{a}$.
This is most easily done by resorting to the standard theory of angular momentum and rotations in terms of spin-1 matrices.
Note that we use these matrices to describe the effect of rotating $\mathbf{a}$ on the numbers
$f(k|\mathbf{a},P,N)$ and that we do not postulate the existence of the spin of a particle.
In our approach, the concept of ``spin'' may be viewed as the result of the interpretation of the mathematical symbols involved,
not necessarily as a postulated, intrinsic property of the particle.

For spin 1, the three spin-1 matrices read~\cite{BALL03}
\begin{eqnarray}
S^x=
\frac{1}{\sqrt{2}}
\left(\begin{array}{rrr}
\phantom{+}0&\phantom{+}1&\phantom{+}0\\
1&0&1\\
0&1&0\\
\end{array}\right)
\quad,\quad
S^y=
\frac{1}{\sqrt{2}}
\left(\begin{array}{rrr}
0&-i&0\\
+i&0&-i\\
0&+i&0\\
\end{array}\right)
\quad,\quad
S^z=
\left(\begin{array}{rrr}
+1&\phantom{+}0&0\\
0&\phantom{+}0&0\\
0&\phantom{+}0&-1\\
\end{array}\right)
,
\label{sec2f4}
\end{eqnarray}
and we immediately see that $\mathbf{\widetilde K}=S^z$.
For completeness, appendix~\ref{APP2} gives a derivation of the well-known result that
a rotation in 3D space which turns a unit vector $\mathbf{u}$ into a unit vector $\mathbf{w}$
corresponds to a rotation in spin-space that changes
the projection of the spin on the direction $\mathbf{u}$
to the projection of the spin on the direction $\mathbf{w}$.
Expressed in a formula, this means that
\begin{eqnarray}
\mathbf{K}({\mathbf{a}})&=&\mathbf{a}\cdot\mathbf{S}
,
\label{sec2f5}
\end{eqnarray}
from which it directly follows that $\mathbf{K}^p({\mathbf{a}})=(\mathbf{a}\cdot\mathbf{S})^p$ for $p=0,1,2$

\section{Matrix representation for filters}{\label{sec2g}}

The next step is consider only those particles which travel along a particular beam $k$
and to construct the corresponding matrices.
As before, it is expedient to start with the case $\mathbf{a}=\mathbf{e}_z$.
Replacing the moments in Eq.~(\ref{sec2a3}) by the powers of $S^z$ we have

\begin{eqnarray}
\mathbf{M}_k(\mathbf{e}_z) &=& \openone-(S^z)^2 + \frac{k}{2} S^z + \frac{k^2}{2}\big[3(S^z)^2 - 2\openone\big]
\nonumber \\
&=&
\left(\begin{array}{ccc}
\frac{k^2+k}{2}&0&0\\
0&1-k^2&0\\
0&0&\frac{k^2-k}{2}\\
\end{array}\right)
=\left\{
\begin{array}{l}
\left(\begin{array}{ccc}
1&0&0\\
0&0&0\\
0&0&0\\
\end{array}\right)\quad,\quad k=+1
\\
\\
\left(\begin{array}{ccc}
0&0&0\\
0&1&0\\
0&0&0\\
\end{array}\right)\quad,\quad k=0
\\
\\
\left(\begin{array}{ccc}
0&0&0\\
0&0&0\\
0&0&1\\
\end{array}\right)\quad,\quad k=-1
\end{array}
\right.
.
\label{sec2g6}
\end{eqnarray}
From Eq.~(\ref{sec2g6}), it follows by inspection that
$\mathbf{M}_k(\mathbf{e}_z) \mathbf{M}_l(\mathbf{e}_z)=\delta_{k,l}\mathbf{M}_k(\mathbf{e}_z)$,
that is the $\mathbf{M}_k(\mathbf{e}_z)$'s are the three mutually orthogonal projectors.
In Appendix~\ref{APP3}, we give a general proof that for a non-degenerate Hermitian
matrix $A$, the projectors onto the eigenspaces of $A$ can be obtained by expanding
a function of the eigenvalues of $A$ in terms of its moments, and then
symbolically replacing each moment by $A$.

As a result of rotating $\mathbf{e}_z$ to $\mathbf{a}$,
$\mathbf{M}_k(\mathbf{e}_z)$ changes into
\begin{eqnarray}
\mathbf{M}_k(\mathbf{a}) &=& 1-(\mathbf{a}\cdot\mathbf{S})^2 + \frac{k}{2} \mathbf{a}\cdot\mathbf{S} +
\frac{k^2}{2}\big[3(\mathbf{a}\cdot\mathbf{S})^2 - 2\openone\big]
.
\label{sec2g7}
\end{eqnarray}
The matrices $\mathbf{M}_k(\mathbf{a})$ represent three mutually orthogonal projectors since
Eq.~(\ref{sec2g7}) follows from Eq.~(\ref{sec2g6}) by a unitary transformation, implying in addition that
$\mathbf{M}_k(\mathbf{a})$ is a Hermitian matrix and $\mathbf{Tr\;}\mathbf{M}_k(\mathbf{a})=1$.
For later use, note that
\begin{eqnarray}
\mathbf{a}\cdot\mathbf{S}=\mathbf{M}_{+1}(\mathbf{a})-\mathbf{M}_{-1}(\mathbf{a})
\quad,\quad
(\mathbf{a}\cdot\mathbf{S})^2=\mathbf{M}_{+1}(\mathbf{a})+\mathbf{M}_{-1}(\mathbf{a})
.
\label{sec2g8}
\end{eqnarray}

\section{Separating the description of the double SG experiment}{\label{sec2h}}

Consistency with the original, non-separated description requires that we have
\begin{eqnarray}
f(k|\mathbf{a},P,N)=\mathbf{Tr\;}\mathbf{F}(P,N)\mathbf{M}_k({\mathbf{a}})=\mathbf{Tr\;}\mathbf{M}_k({\mathbf{a}})\mathbf{F}(P,N)
=\mathbf{Tr\;}\mathbf{M}_k({\mathbf{a}})\mathbf{F}(P,N)\mathbf{M}_k({\mathbf{a}})
,
\label{sec2h0}
\end{eqnarray}
where we have used the invariance of the trace under cyclic permutation of the matrices
and the fact that $\mathbf{M}_k(\mathbf{a})$ is a projector to write down three equivalent forms.
Note that Born's rule~\cite{BORN26} {\bf postulates} Eq.~(\ref{sec2h0}) whereas in
the approach taken in this paper, Eq.~(\ref{sec2h0}) is obtained
by selecting, from the many different ways of representing the frequencies of events $f(k|\mathbf{a},P,N)$
and the averages computed from them, the one that yields a description which is separated in parts.

The next step is to extend the separated description of the SG experiment
in terms of $\mathbf{F}(P,N)$ and $\mathbf{M}_k({\mathbf{a}})$ to the double SG experiment.

\begin{center}
\framebox{
\parbox[t]{0.95\hsize}{%
As all SG magnets are assumed to be identical, consistency demands that their description should be
the same, that is the filtering property of SG2, SG3 and SG4
should be described by $\mathbf{M}_l({\mathbf{b}})$.
}}
\end{center}

The question now is how to generalize Eq.~(\ref{sec2h0}) to yield $f(k,l|\mathbf{a},\mathbf{b},P,N)$.
As $\mathbf{F}(P,N)$ completely characterizes the particles leaving the source
and $\mathbf{M}_k({\mathbf{a}})$ determines the number of particles
that exit SG1 through beam $k$,
we could try to interpret the matrix product $\mathbf{M}_k({\mathbf{a}})\mathbf{F}(P,N)$ as
a ``new source'' emitting particles along beam $k$ towards the second stage of SG magnets.
For the sake of argument, let us interpret $\mathbf{M}_k({\mathbf{a}})\mathbf{F}(P,N)$
as representing the source $\mathbf{F}(P,N)$ emitting particles followed by beam selection through $\mathbf{M}_k({\mathbf{a}})$.
Then, we would read $\mathbf{M}_l({\mathbf{b}})\mathbf{M}_k({\mathbf{a}})\mathbf{F}(P,N)$
as the source $\mathbf{F}(P,N)$ emitting particles, beam selection by $\mathbf{M}_k({\mathbf{a}})$,
followed by beam selection through $\mathbf{M}_l({\mathbf{b}})$.
Although this may sound reasonable, this interpretation leads to inconsistencies
because the only thing that matters is
the result that we obtain by calculating the trace of the matrix product.
Indeed, as $\mathbf{Tr\;}\mathbf{M}_l({\mathbf{b}})\mathbf{M}_k({\mathbf{a}})\mathbf{F}(P,N)=
\mathbf{Tr\;}\mathbf{M}_k({\mathbf{a}})\mathbf{F}(P,N)\mathbf{M}_l({\mathbf{b}})$
we would read the latter as ``a source $\mathbf{M}_l({\mathbf{b}})$ emits particles, ...'', which clearly makes no sense.
Using this line of reasoning, it is not too difficult to convince oneself that
the only expression that has a contradiction-free meaning is the last one of Eq.~(\ref{sec2h0}).
In words, we say that the results of filtering by $\mathbf{M}_k({\mathbf{a}})$
is to produce a fictitious source in beam $k$ which is described by the matrix
$\mathbf{M}_k({\mathbf{a}})\mathbf{F}(P,N)\mathbf{M}_k({\mathbf{a}})$.
The latter is also the only form which satisfies the requirement that the matrix describing
the source must be Hermitian (see Section 9).
Consistency with the earlier expression then requires that
\begin{eqnarray}
f(k,l|\mathbf{a},\mathbf{b},P,N)=\mathbf{Tr\;}\mathbf{M}_l({\mathbf{b}})\mathbf{M}_k({\mathbf{a}})\mathbf{F}(P,N)\mathbf{M}_k({\mathbf{a}})\mathbf{M}_l({\mathbf{b}})
.
\label{sec2h1}
\end{eqnarray}

A direct consequence of Eq.~(\ref{sec2h1}) is that
\begin{eqnarray}
f(k|\mathbf{a},P,N)&=& \sum_{l=+1,0,-1} f(k,l|\mathbf{a},\mathbf{b},P,N)
,
\label{sec2h2}
\end{eqnarray}
which expresses the fact that in the double SG experiment,
the frequencies of outcomes after the first SG magnet (SG1) are a function of $\mathbf{a}$ only,
a direct consequence of the application of \POS.

Although Eq.~(\ref{sec2h1}) can be simplified to
$f(k,l|\mathbf{a},\mathbf{b},P,N)=\mathbf{Tr\;}\mathbf{M}_l({\mathbf{b}})\mathbf{M}_k({\mathbf{a}})\mathbf{F}(P,N)\mathbf{M}_k({\mathbf{a}})$,
Eqs.~(\ref{sec2h0}) and Eq.~(\ref{sec2h1}) make it clear how the approach generalizes to three, four, ... layers of SG magnets.

\begin{center}
\framebox{
\parbox[t]{0.95\hsize}{%
If we interpret $\mathbf{F}(P,N)$ as the $3\times3$ density matrix $\rho$ which
characterizes the state of a quantum system, then Eqs.~(\ref{sec2h0}) and (\ref{sec2h1}) are exactly
the same as those postulated in quantum theory~\cite{BALL03}.
}}
\end{center}

\section{Illustrative example}{\label{sec2i}}

Up to this point, the magnetic properties of particles before they interact with the first SG magnet,
represented by the symbol $P$, did not play any role (apart from the assumption that the magnetic field affects the particles).
As an example we consider the case in which $P$ corresponds to the matrix
\begin{eqnarray}
\mathbf{F}(P,N)=
\frac{1}{3}
\left(\begin{array}{ccc}
1&0&0\\
0&1&0\\
0&0&1\\
\end{array}\right)
,
\label{sec2i0}
\end{eqnarray}
and ask ourselves what we can learn about the magnetic properties of the particles
by performing the double SG experiment.

Performing the matrix multiplications and calculating traces yields
\begin{eqnarray}
f(k|\mathbf{a},P,N)&=&\mathbf{Tr\;}\mathbf{M}_k({\mathbf{a}})\mathbf{F}(P,N)\mathbf{M}_k({\mathbf{a}})=\frac{1}{3}
,
\nonumber\\
\\
\langle k^p \rangle &=& \mathbf{Tr\;}\mathbf{F}(P,N)\mathbf{K}^p({\mathbf{a}})
= \mathbf{Tr\;}\mathbf{F}(P,N)({\mathbf{a}}\cdot\mathbf{S})^p
\nonumber\\
&=&
\left\{
\begin{array}{l}
\mathbf{Tr\;}\mathbf{F}(P,N)=1 \quad,\quad p=0 \\ \\
\mathbf{Tr\;}\mathbf{F}(P,N)(\mathbf{M}_{+1}(\mathbf{a})-\mathbf{M}_{-1}(\mathbf{a}))=0 \quad,\quad p=1 \\ \\
\mathbf{Tr\;}\mathbf{F}(P,N)(\mathbf{M}_{+1}(\mathbf{a})+\mathbf{M}_{-1}(\mathbf{a}))=\frac{2}{3} \quad,\quad p=2 \\
\end{array}
\right.
.
\label{sec2i1}
\end{eqnarray}
and
\begin{eqnarray}
f(k,l|\mathbf{a},\mathbf{b},P,N)
&=&\mathbf{Tr\;}\mathbf{M}_l({\mathbf{b}})\mathbf{M}_k({\mathbf{a}})\mathbf{F}(P,N)\mathbf{M}_k({\mathbf{a}})\mathbf{M}_l({\mathbf{b}})
\nonumber\\
&=&
\left\{
\begin{array}{l}
\frac{1}{12}(1+\mathbf{a}\cdot\mathbf{b})^2 \quad,\quad k=l=+1,-1 \\ \\
\frac{1}{3}(\mathbf{a}\cdot\mathbf{b})^2 \quad,\quad k=l=0 \\ \\
\frac{1}{12}(1-\mathbf{a}\cdot\mathbf{b})^2 \quad,\quad (k,l)=(+1,-1),(-1,+1) \\ \\
\frac{1}{6}(1-(\mathbf{a}\cdot\mathbf{b})^2) \quad,\quad (k,l)=(+1,0),(-1,0),(0,+1),(0,-1) \\
\end{array}
\right.
.
\nonumber\\
\label{sec2i2}
\end{eqnarray}

From Eqs.~(\ref{sec2i1}) it is clear that the description of the counts in beams $k=+1,0,-1$ does not depend
on ${\mathbf{a}}$. In other words, the choice Eq.~(\ref{sec2i0}) of $\mathbf{F}(P,N)$ describes a situation that
is invariant under rotations of $\mathbf{a}$.
Similarly, Eq.~(\ref{sec2i2}) shows that the dependence of the outcomes on
the directions $\mathbf{a}$ and $\mathbf{b}$ of the respective magnetic fields only
enters through the angle between the two vectors $\mathbf{a}$ and $\mathbf{b}$.
On the other hand, there is a-priori no reason why
$f(k,l|\mathbf{a},\mathbf{b},P,N)$
should depend on $\mathbf{a}\cdot\mathbf{b}$ only.
The dependence on $\mathbf{a}\cdot\mathbf{b}$ is a direct consequence of the choice Eq.~(\ref{sec2i0}) of $\mathbf{F}(P,N)$
and the desire to separate the description into independent descriptions of parts.
From Eq.~(\ref{sec2i2}) it is clear that $p(k,l|\mathbf{a},\mathbf{a})=\delta_{k,l}/3$.
Therefore, this model of the SG magnet functions as an ideal filtering device,
meaning that it is possible to assign a definite magnetic moment to the particle.

The reasoning that led to the general form Eq.~(\ref{sec2h1}) and to the example Eq.~(\ref{sec2i2})
does not predict but rather restricts the functional dependence
of the frequencies $f(k,l|\mathbf{a},\mathbf{b},P,N)$ on $\mathbf{a}$ and $\mathbf{b}$.
For instance, and only for the sake of argument, if we replace in Eq.~(\ref{sec2i2})
$\mathbf{a}\cdot\mathbf{b}$ by $(\mathbf{a}\cdot\mathbf{b})^4$, the
resulting expression for $f(k,l|\mathbf{a},\mathbf{b},P,N)$ are valid
frequencies that might be realized in a (computer) experiment but do not admit a description
in terms of quantum theory.
Indeed, such expressions {\bf cannot} be obtained from the quantum theoretical considerations
because the projectors Eq.~(\ref{sec2g7}) are quadratic functions of $\mathbf{a}\cdot\mathbf{S}$
(or of $\mathbf{b}\cdot\mathbf{S}$).
In other words, we have $\mathrm{SOC}\models \mathrm{QT}$.
For an explicit example in the context of the EPRB experiment, see Ref.~\cite{RAED18a}.

We summarize these findings as follows:
\begin{center}
\framebox{
\parbox[t]{0.95\hsize}{%
\begin{enumerate}
\item
There exist physically realizable processes (e.g. computer simulations) that produce data
which do not allow for a separation of the form Eq.~(\ref{sec2e3a}).
\item
As explained above and demonstrated explicitly in Ref.~\cite{RAED18a},
there also exist physically realizable processes
that produce data which allow for a separation of the form Eq.~(\ref{sec2e3a})
but are outside the scope of what standard quantum theory can possibly describe.
\item
Therefore, the quantum formalism describes a proper (strict) subset
of a class of experiments for which \POS\ holds, i.e, $\mathrm{SOC}\models \mathrm{QT}$.
\end{enumerate}
}}
\end{center}

\section{General description of the source}{\label{sec2j}}

In Sec.~\ref{sec2i}, we considered the special and also simple case in which
the source is described by the matrix $\mathbf{F}(P,N)=\openone/3$.
The most general description of the magnetic properties of the
particles before they enter the magnetic field maintained by the first SG magnet
can be constructed as follows.
First, we choose a complete basis for the linear space of $3\times3$ matrices which is orthonormal with respect
to the inner product $(A,B)\equiv \mathbf{Tr\;} A^\dagger B$.
For instance, one possible choice is
\begin{eqnarray}
{\cal B}&=&\bigg(\mathbf{B}_0,\ldots,\mathbf{B}_8\bigg)=
\bigg(
\frac{\openone}{\sqrt{3}},
\frac{S^x}{\sqrt{2}},
\frac{S^y}{\sqrt{2}},
\frac{S^z}{\sqrt{2}},
-\sqrt{\frac{2}{3}}\openone +\sqrt{\frac{3}{2}}(S^x)^2,
-\sqrt{2}\openone +\frac{(S^x)^2}{\sqrt{2}} +\sqrt{2}(S^z)^2,
\nonumber \\
&&\hbox to 3cm{}
\frac{S^xS^y + S^yS^x}{\sqrt{2}},
\frac{S^xS^z + S^zS^x}{\sqrt{2}},
\frac{S^yS^z + S^zS^y}{\sqrt{2}}
\bigg)
,
\label{sec2j0}
\end{eqnarray}
is such a basis. We have $(B_i,B_j)=\delta_{i,j}$ for $i,j=0,\ldots,8$ and
in addition, we have $\mathbf{Tr\;} \mathbf{B}_0=\sqrt{3}$ and $\mathbf{Tr\;} \mathbf{B}_i=0$ for $i=1,\ldots,8$.

With the help of this basis, we can write down the most general expression for $\mathbf{F}(P,N)$ as
\begin{eqnarray}
\mathbf{F}(P,N)&=&\sum_{i=0}^8 f_i \mathbf{B}_i
,
\label{sec2j1}
\end{eqnarray}
where the expansion coefficients $f_i$'s can, in principle, be arbitrary complex-valued numbers.
Imposing the restriction that $\mathbf{Tr\;} \mathbf{F}(P,N)=1$ enforces $f_0=1/\sqrt{3}$.
The other coefficients can only be determined from the observed data.
Using expansion Eq.~(\ref{sec2j1}) we find
\begin{eqnarray}
\langle k \rangle = \mathbf{Tr\;}\mathbf{F}(P,N)\;{\mathbf{a}}\cdot\mathbf{S}
= \sqrt{2}(a_x f_1 +a_y f_2 + a_z f_3)
,
\label{sec2ej2}
\end{eqnarray}
and
\begin{eqnarray}
\langle k^2 \rangle &=& \mathbf{Tr\;}\mathbf{F}(P,N)\;({\mathbf{a}}\cdot\mathbf{S})^2
\nonumber \\
&=& \frac{2}{\sqrt{3}}
+\sqrt{\frac{2}{3}} f_4 a_x^2
-(\frac{f_4}{\sqrt{6}}+\frac{f_5}{\sqrt{2}})a_y^2
-(\frac{f_4}{\sqrt{6}}-\frac{f_5}{\sqrt{2}})a_z^2
+\sqrt{2}(a_xa_y f_6 + a_xa_zf_7+a_ya_z f_8)
.
\label{sec2ej3}
\end{eqnarray}
As Eqs.~(\ref{sec2ej2}) and (\ref{sec2ej3}) are linear in the unknown $f_i$'s,
the latter can be found by solving the two linear sets of equations
obtained by repeating the experiment with five different values of ${\mathbf{a}}$.
For each of these five values of ${\mathbf{a}}$, the experiment
yields values of $\langle k \rangle$ and $\langle k^2 \rangle$.
Three of such values of $\langle k \rangle$ suffice to determine $f_1$, $f_2$, and $f_3$.
The five values of $\langle k^2 \rangle$ allow us to solve for $f_4$, $f_5$, $f_6$, $f_7$, and $f_8$.
The left-hand-sides of Eqs.~(\ref{sec2ej2}) and (\ref{sec2ej3}), being obtained by counting,
are necessarily real-valued numbers.
As Eqs.~(\ref{sec2ej2}) and (\ref{sec2ej3}) hold for any choice of ${\mathbf{a}}$,
it follows immediately that all the $f_i$'s must be real-valued numbers too.
By choice, the basis vectors are Hermitian matrices.
Therefore, requiring the description of the data to be separable automatically
enforces the matrix $\mathbf{F}(P,N)$ to be Hermitian.
Furthermore, $\mathbf{Tr\;}\mathbf{M}_k({\mathbf{a}})\mathbf{F}(P,N)\mathbf{M}_k({\mathbf{a}})$
corresponds to the counts in beam $k\in{\cal E}$ and must therefore be a non-negative number
for all choices of ${\mathbf{a}}$.
As $\mathbf{M}_k({\mathbf{a}})$ is a projector on the $k$th eigenstate
$\widehat{\mathbf{a}}_k$ of ${\mathbf{a}}\cdot{\mathbf{S}}$,
i.e. $\mathbf{M}_k({\mathbf{a}})=\widehat{\mathbf{a}}_k\widehat{\mathbf{a}}_k^\mathrm{T}$,
we have
$\mathbf{Tr\;}\mathbf{M}_k({\mathbf{a}})\mathbf{F}(P,N)\mathbf{M}_k({\mathbf{a}})=
\mathbf{Tr\;}\widehat{\mathbf{a}}_k\widehat{\mathbf{a}}_k^\mathrm{T}\mathbf{F}(P,N)\widehat{\mathbf{a}}_k\widehat{\mathbf{a}}_k^\mathrm{T}
=\widehat{\mathbf{a}}_k^{\mathrm{T}}\cdot \mathbf{F}(P,N) \cdot \widehat{\mathbf{a}}_k\ge0$
for all unit vectors ${\mathbf{a}}$, implying that the matrix $\mathbf{F}(P,N)$ is positive semidefinite.
Obviously, $\mathbf{F}(P,N)$ has all the properties of the
density matrix $\rho$, which in quantum theory, is postulated to be the mathematical representation
of the state of the system~\cite{BALL03}.

\section{Relation to Heisenberg matrix mechanics}\label{sec9}

From Sec.~\ref{sec2}, it is clear that the use of matrix algebra is key to construct,
starting from the notion of individual events, the mathematical structure of quantum theory.
Matrix algebra also played a key role in the early development of quantum theory~\cite{NEUM55,WEIN03},
so let us briefly review the essential elements of Heisenberg's matrix mechanics~\cite{HEIS25}.

Consider a classical mechanical, one-particle system characterized by the Hamiltonian $H(p,q)$
where $p$ and $q$ are the momentum and position of the particle, respectively.
According to Heisenberg's recipe, we seek for some representation of $p$ and $q$ in
terms of two matrices $\widehat p$ and $\widehat q$ such that $[\widehat q,\widehat p]=i\hbar\openone$
and that the matrix $H(\widehat p,\widehat q)$ becomes diagonal~\cite{NEUM55,WEIN03}.
The diagonal elements of this matrix are the eigenvalues of the system
and the matrix elements of $\widehat q$ can be used to compute transition
rates between the eigenstates of the system~\cite{NEUM55,WEIN03}.
In Heisenberg's construction, the two-indexed objects (that is, the matrices)
appear because of Heisenberg's assumption that, rather than the atomic states themselves,
only {\sl transitions} between atomic states (that is, pairs of initial and final states) are observable.
Note that the matrices $\widehat p$
and $\widehat q$ cannot be finite dimensional because that would be in conflict
with the statement that the trace of the commutator of two finite-dimensional
matrices is zero~\cite{SHOD36,ALBE57}.

As is well-known, Heisenberg's matrix mechanics can be derived from Schr\"odinger's wave mechanics~\cite{SCHR26c,WEIN03}.
Both approaches postulate a mathematical structure that leads to the desirable features such as
discrete energy levels. On this level of description, there is no connection to individual detection events.
This comes in through Born's rule~\cite{BORN26} which postulates that the probability to observe
a particle at a point $q$ is given by the modulus squared of the wave function at this point.
The chain of reasoning in this case is the one depicted in Fig.~\ref{nopostulates}(a) which
conceptually is very different from Fig.~\ref{nopostulates}(c).
Therefore, except for the use of the machinery of matrix calculus itself,
there is no direct relation between Heisenberg's matrix mechanics
and the approach pursued in this paper.

\section{Parameter dependence}\label{sec4}

Next, we consider a source whose characteristics change as a function of a parameter $\lambda$.
For each value $\lambda$ of interest, we detect $N$ particles and construct
the data set ${\cal D}(\lambda)$, as explained in Sec.~\ref{sec2a}.
As before, from this data we compute
$f(k,l|\mathbf{a},\mathbf{b},P,N,\lambda)$ and  $f(k|\mathbf{a},P,N,\lambda)$.
According to \POS, we have
\begin{eqnarray}
\langle k^p \rangle_\lambda = \mathbf{Tr\;}\mathbf{F}(P,N,\lambda)\mathbf{K}^p({\mathbf{a}})
\quad,\quad p=0,1,2
,
\label{sec4a1}
\end{eqnarray}
where the notation $\langle k^p \rangle_\lambda$ is used to make explicit that
the first and second moments depend on $\lambda$.
From $\mathbf{Tr\;}\mathbf{F}(P,N,\lambda)=1$, it follows immediately that
\begin{eqnarray}
\mathbf{Tr\;}\frac{\partial^n \mathbf{F}(P,N,\lambda)}{\partial\lambda^n} = 0 \quad,\quad n>0
,
\label{sec4a2}
\end{eqnarray}
meaning that all the derivatives of $\mathbf{F}(P,N,\lambda)$ with respect
to $\lambda$ are traceless matrices and it is understood that these derivatives are well-defined.
As a traceless matrix is the commutator of two matrices~\cite{SHOD36,ALBE57}, we may write
\begin{eqnarray}
\frac{\partial \mathbf{F}(P,N,\lambda)}{\partial\lambda} = [Y(\lambda),Z(\lambda)]
,
\label{sec4a3}
\end{eqnarray}
where $Y(\lambda)$ and $Z(\lambda)$ are matrices of the same dimension as $\mathbf{F}(P,N,\lambda)$.

On the other hand, $\mathbf{F}(P,N,\lambda)$ is a Hermitian (non-negative definite)  matrix
and can therefore be written as $\mathbf{F}(P,N,\lambda)=U^\dagger(\lambda)D(\lambda)U(\lambda)$
where $D(\lambda)$ are the non-negative eigenvalues of $\mathbf{F}(P,N,\lambda)$ and
$U(\lambda)$ is the unitary transformation which diagonalizes $\mathbf{F}(P,N,\lambda)$
(here and in the remainder of this section, we write $U(\lambda)=U(P,N,\lambda)$, etc.~in
order to simplify the notation).
Using
$\partial \left(U^\dagger(\lambda)U(\lambda)\right)/\partial \lambda=
\left(\partial U^\dagger(\lambda)/\partial \lambda\right)U(\lambda)+
U^\dagger(\lambda)\left(\partial U(\lambda)/\partial \lambda\right)=0$,
we have
\begin{eqnarray}
\frac{\partial \mathbf{F}(P,N,\lambda)}{\partial\lambda}=
\bigg[\mathbf{F}(P,N,\lambda),U^\dagger(\lambda)\frac{\partial U(\lambda)}{\partial \lambda}\bigg]
+
U^\dagger(\lambda)\frac{\partial D(\lambda)}{\partial \lambda}U(\lambda)
.
\label{sec4a4a}
\end{eqnarray}
In the following, we examine the case where all the eigenvalues of $\mathbf{F}(P,N,\lambda)$ are independent of $\lambda$.
In this case,  $\partial D(\lambda)/\partial \lambda=0$,
and comparing Eq.~(\ref{sec4a3}) and Eq.~(\ref{sec4a4a}) shows that,
up to irrelevant additive terms and factors, $Y(\lambda)=\mathbf{F}(P,N,\lambda)$
and $Z(\lambda)=U^\dagger(\lambda)(\partial U(\lambda)/\partial \lambda)$.
As $Z^\dagger(\lambda)=-Z(\lambda)$ we may write $Z(\lambda)\equiv iH(\lambda)$
where $H(\lambda)$ is a Hermitian matrix.
%
%
%
From Eq.~(\ref{sec4a3}) it then follows that
\begin{eqnarray}
\frac{\partial \mathbf{F}(P,N,\lambda)}{\partial\lambda} = i[\mathbf{F}(P,N,\lambda),H(\lambda)]
.
\label{sec4a5}
\end{eqnarray}
The formal solution of Eq.~(\ref{sec4a5}) reads
\begin{eqnarray}
\mathbf{F}(P,N,\lambda) = V(\lambda)\mathbf{F}(P,N,0)V^\dagger(\lambda)
,
\label{sec4a6a}
\end{eqnarray}
where the unitary matrix $V(\lambda)$ is the solution of
\begin{eqnarray}
i \frac{\partial V(\lambda)}{\partial\lambda} = H(\lambda)V(\lambda)\quad,\quad V(0)=\openone
,
\label{sec4a7a}
\end{eqnarray}
which has the structure of the time-dependent Schr\"odinger equation.
In other words, if we restrict ourselves to the class of data for which the eigenvalues of $\mathbf{F}(P,N,\lambda)$
do not depend on $\lambda$, the parameter dependence of $\mathbf{F}(P,N,\lambda)$
is determined by an equation that is reminiscent of the time-evolution equation
of a closed quantum system (in general, the time-evolution of open quantum system
cannot be described in terms of a unitary matrix~\cite{BREU02}).
Clearly, the restriction to cases for which the eigenvalues of $\mathbf{F}(P,N,\lambda)$
do not depend on $\lambda$ is yet another indication that $\mathrm{SOC} \models \mathrm{QT}$.

Note that Eq.~(\ref{sec4a6a}) is consistent with the assumption
that \POS\ holds for all $\lambda$.
Indeed, using Eq.~(\ref{sec4a6a}) we have
$\langle k^p \rangle_\lambda = \mathbf{Tr\;}\mathbf{F}(P,N,0)
V^\dagger(\lambda)\mathbf{K}^p({\mathbf{a}})V(\lambda)$
for $p=0,1,2$, which has the form that we expect from the application of \POS.
Whether there are more general solutions of Eqs.~(\ref{sec4a3}) and (\ref{sec4a4a})
that are compatible with \POS\ is an open question.

\subsection{von Neumann and Schr\"odinger equation}\label{sec4a}

After proper identification of the symbols and introducing units of time and energy, Eq.~(\ref{sec4a5})
is nothing else but the von Neumann equation
\begin{eqnarray}
i\hbar\frac{\partial \rho(t)}{\partial t} = [H(t),\rho(t)]
,
\label{sec4a6}
\end{eqnarray}
for the density matrix~\cite{BREU02}.
Note that adding to $H(t)$
the matrix $c(t)\openone$, $c(t)$ being a complex number, does not change Eq.~(\ref{sec4a6}).
Traditionally, the von Neumann Eq.~(\ref{sec4a6}) is obtained from
the Schr\"odinger equation of a pure state after introducing the concept of a
weighted mixture of pure states~\cite{NEUM55,BREU02,BALL03}.
Conversely, the Schr\"odinger equation follows from the
von Neumann Eq.~(\ref{sec4a6}) if we assume that $\rho(t)$ takes the form of a
pure state $|\Psi(t)\rangle$, i.e. $\rho(t)=|\Psi(t)\rangle\langle\Psi(t)|$.
In this case, Eq.~(\ref{sec4a6}) reads
\begin{eqnarray}
i\hbar
\frac{\partial |\Psi(t)\rangle}{\partial t} \langle\Psi(t)|
+
i\hbar
|\Psi(t)\rangle\frac{\partial \langle\Psi(t)|}{\partial t}
=
H(t)|\Psi(t)\rangle\langle\Psi(t)| - |\Psi(t)\rangle\langle\Psi(t)|H(t)
,
\label{sec4a7}
\end{eqnarray}
which is (up to an irrelevant phase-shift matrix $\exp(i c(t) \openone)$) equivalent to the time-dependent Schr\"odinger equation
\begin{eqnarray}
i\hbar \frac{\partial }{\partial t}|\Psi(t)\rangle =H(t)|\Psi(t)\rangle
.
\label{sec4a8}
\end{eqnarray}

From Eq.~(\ref{sec4a8}), it follows that the matrix $H(t)$, playing the role
of the time-dependent Hamiltonian,
is the generator of infinitesimal time displacements of a vector $|\Psi(t)\rangle$.
Returning to the specific example of the SG experiment with three different outcomes, this matrix takes the general form
\begin{eqnarray}
H(t)&=&\sum_{i=1}^8 h_i(t) \mathbf{B}_i
,
\label{sec4a9}
\end{eqnarray}
where the eight expansion coefficients $h_i(t)$ are real numbers
the $\mathbf{B}_i$'s are defined by Eq.~(\ref{sec2j0}),
and we have dropped the term with $\mathbf{B}_0=\openone/\sqrt{3}$ because adding such a term to $H(t)$ does not change Eq.~(\ref{sec4a6}).
We repeat that our treatment trivially generalizes to experiments with any number of different outcomes.

It may be worthwhile to mention here that many applications of quantum physics to e.g. condensed matter problems
de facto start from a representation such as Eq.~(\ref{sec4a9}).
For instance, the description of electron paramagnetic resonance spectra
usually starts from a single-spin Hamiltonian such as Eq.~(\ref{sec4a9})
that contains Zeeman terms, the interaction with the crystal field (for $S>1/2$) and other interactions of
the magnetic moment with its environment~\cite{ABRA70}.
In practice, the parameters which specify the strength of the various contributions
to the Hamiltonian are obtained by fitting the model to experimental data.

Originally, the Schr\"odinger equation for a particle in a potential was formulated in continuum space~\cite{SCHR26a}.
In contrast, the construction of the quantum theoretical framework presented in this paper builds on
data that is represented by a finite number of different kinds of events (e.g. $k=-1,0,1$), i.e. by finite-dimensional matrices.
The transition from the finite-dimensional to the infinite-dimensional (continuum) case
is very nicely and extensively explained in the Feynman lectures~\cite{FEYN65},
and will therefore not be repeated here.

Historically, classical Hamiltonian mechanics served as the starting point
for formulating the corresponding quantum mechanical problem, see e.g. our discussion of Heisenberg's matrix mechanics~\cite{HEIS25}
and Schr\"odinger's first derivation of his equation~\cite{SCHR26a}.
However, for particles moving in continuum space, the symmetries of space-time very much determine
the form of the Hamiltonian in terms of the operators that correspond to momentum,
angular momentum, potentials etc.~\cite{JORD69,BALL03,WEIN03},
without recourse to classical mechanics.
Therefore, completing the present construction with the part giving physical content
to the description is not a real issue.

In summary, we have shown that a description of the time-dependent
data set ${\cal D}(t)$ does not require us to postulate Eqs.~(\ref{sec4a6}) or (\ref{sec4a8}).
We emphasize that in both the time-dependent and time-independent case,
we have $\mathrm{SOC} \models \mathrm{QT}$, i.e., \POS\ allows for equations that
are not compatible with quantum theory.
However, much if not all of the machinery of quantum theory for a single particle
follows from \POS, a straightforward application of matrix algebra, and
the symmetries of the space-time continuum.
There is no need to introduce postulates about ``wave functions'', ``observables'', ``quantization rules'',
``Born's rule'', and the like.

It remains to be shown how the same ideas extend to the case
that the data consists of tuples $(k_1,k_2,\ldots)$ instead of a single item $k$.
This problem is the subject of the next section where we show, by means
of a concrete example of a two-particle problem, how the
direct-product structure of vector spaces and matrices emerges from application of \POS\ in a most natural manner.

\begin{figure}[h]
\centering
\includegraphics[width=0.8\hsize]{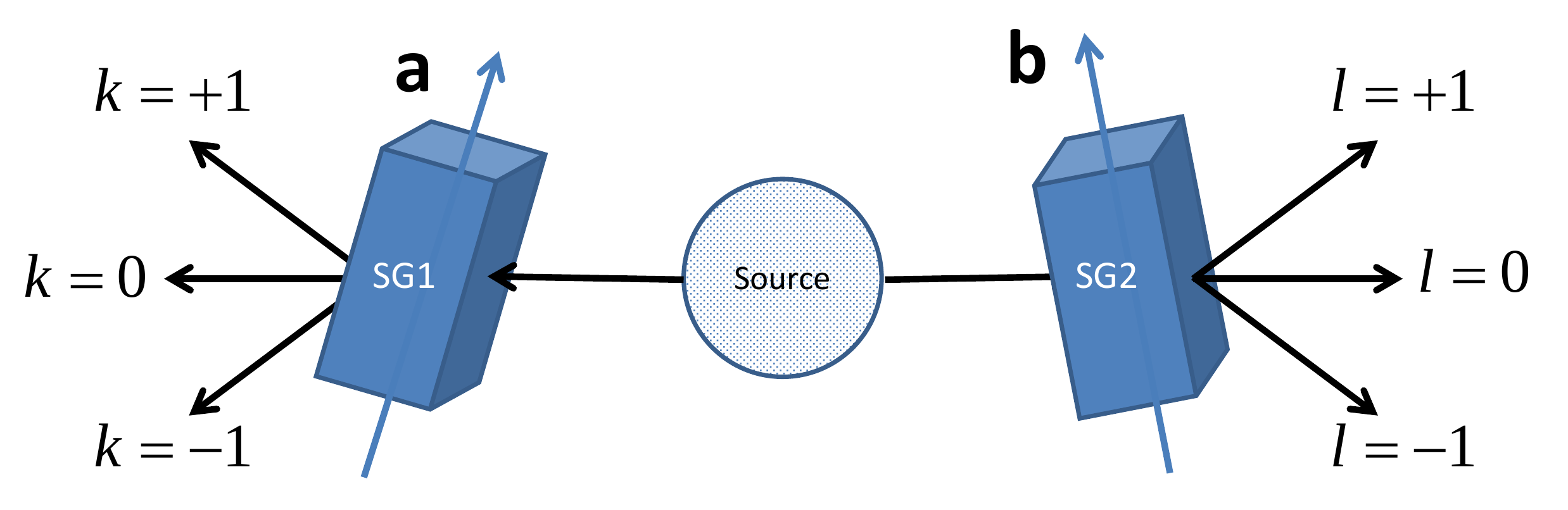}  
\caption{(Color online)
Layout of the EPRB thought experiment with pairs of magnetic particles.
The inhomogeneous magnetic field created by Stern-Gerlach magnet SG1 (SG2)
is characterized by the unit vector $\mathbf{a}$ ($\mathbf{b}$).
A particle passing through SG1 (SG2) appears in one of the beams labeled by
$k=+1,0,-1$ ($l=+1,0,-1$).
}
\label{fig1}
\end{figure}

\section{Einstein-Podolsky-Rosen-Bohm experiment}\label{sec3}

This section is not meant to
contribute to the Einstein-Bohr debate~\cite{HESS15}, related to
a Gedanken-experiment suggested by Einstein-Podolsky-Rosen~\cite{EPR35} and modified by Bohm~\cite{BOHM51}.
Its purpose is to demonstrate how the quantum theoretical description
follows from the application of \POS, the requirement of consistency with the
description of the single- and double SG experiment developed above,
and space-time continuum symmetries,
without resorting to one of the postulates of quantum theory.
Most importantly, this section shows, by means of the simplest example of a two-particle system,
how the direct-product-of-Hilbert-spaces structure, which is characteristic for many-body quantum theory,
emerges from the application of \POS.

The layout and data gathering procedure of
the EPRB thought experiment that we consider is illustrated by Fig.~\ref{fig1}.
The experiment produces a data set
\begin{equation}
{\cal D}=\big\{ (k_n,l_n)\;|\; k_n,l_n\in{+1,0,-1}\;;\; n=1,\ldots ,N\big\}
.
\label{sec3a0}
\end{equation}
for each pair of settings $(\mathbf{a},\mathbf{b})$.
From this data set we can compute the relative frequency of an event $(k,l)$
\begin{eqnarray}
f(k,l|\mathbf{a},\mathbf{b},P,N)&=&\frac{1}{N}\sum_{n=1}^N \delta_{k,k_n}\delta_{l,l_n}
.
\label{sec3a1}
\end{eqnarray}
The notation used and the structure are the same as before, see Eq.~(\ref{sec2a2a}).
However, the meaning of $k$ and $l$ are quite different from that in Sec.~\ref{sec2a}
because $k$ and $l$ refer to the detection of two particles, not of one.
Therefore, we cannot proceed in a sequential manner by considering only one SG magnet and add the second one later.

As in Sec.~\ref{sec2a}, we only consider the simplest case where we
discard all knowledge about the events that is not contained in $f(k,l|\mathbf{a},\mathbf{b},P,N)$.
Here and in the remainder of this section, the symbol $P$ indicates a conditional dependence
on the properties of both particles.

Adopting the same reasoning as for the SG experiment, it follows that we cannot separate the description
of the EPRB data in different parts if we stick to a representation in terms of vectors.
Therefore, we simply repeat the steps that led to the matrix
representation in Eq.~(\ref{sec2e3a}) and start by writing the observations and relative frequencies
as $9\times9$ diagonal matrices
\begin{eqnarray}
\left(\mathbf{\widetilde K}_1\right)_{[k,l],[k,l]}=k
\quad,\quad
\left(\mathbf{\widetilde K}_2\right)_{[k,l],[k,l]}=l
\quad\mathrm{and}\quad
\left(\mathbf{\widetilde F}(\mathbf{a},\mathbf{b},P,N)\right)_{[k,l],[k,l]}=f(k,l|\mathbf{a},\mathbf{b},P,N)
\quad,\quad (k,l)\in{\cal E}^2
,
\label{sec3a2}
\end{eqnarray}
where we have introduced the notation $A_{[k,l],[k',l']}=A_{i(k,l),i(k',l')}$
for the matrix elements of $A$ and the function $i(k,l)=2-k+3(1-l)$ is only there to map $(k,l)\in{\cal E}^2$
onto the standard matrix indices which run from 1 to 9.
The reason for introducing two different matrices
$\mathbf{\widetilde K}_1$
and
$\mathbf{\widetilde K}_2$
representing the observations is that in an EPRB experiment, it is obviously necessary
to distinguish between detector clicks in the left (subscript 1) and right (subscript 2) wing of the experiment, see Fig.~\ref{fig1}.

Separating the descriptions of the source and observation stations
means that we search for $9\times9$ matrices $\mathbf{F}(P,N)$,
$\mathbf{K}_1({\mathbf{a}})$, and
$\mathbf{K}_2({\mathbf{b}})$ such that
\begin{eqnarray}
\langle k^pl^q \rangle=\mathbf{Tr\;}\widetilde{\mathbf{F}}(P,N,\mathbf{a},\mathbf{b})
\widetilde{\mathbf{K}}_1^p \widetilde{\mathbf{K}}_2^q
=\mathbf{Tr\;}\mathbf{F}(P,N){\mathbf{K}}_1^p(\mathbf{a}){\mathbf{K}}_2^q(\mathbf{b})
\quad,\quad p,q=0,1,2
.
\label{sec3a3}
\end{eqnarray}
Consistency of the description with the one of the (double) SG experiments
dictates that if $q=0$ ($p=0$), we must have
\begin{equation}
\mathbf{K}_1(\mathbf{a})=(\mathbf{a}\cdot\mathbf{S}_1)\otimes\openone
\quad,\quad
\mathbf{K}_2(\mathbf{b})=\openone\otimes(\mathbf{b}\cdot\mathbf{S}_2)
,
\label{sec3a4}
\end{equation}
where $\otimes$ denotes the Kronecker product and $\openone$ is the $3\times3$ unit matrix.
Similarly, the expressions for the projections are given by
\begin{eqnarray}
f(k,l|\mathbf{a},\mathbf{b},P,N)
=\mathbf{Tr\;}
 {\mathbf{M}}_l^{(2)}(\mathbf{b}){\mathbf{M}}_k^{(1)}(\mathbf{a})
\mathbf{F}(P,N)
{\mathbf{M}}_k^{(1)}(\mathbf{a}) {\mathbf{M}}_l^{(2)}(\mathbf{b})
=\mathbf{Tr\;}
\mathbf{F}(P,N)
{\mathbf{M}}_k^{(1)}(\mathbf{a}) {\mathbf{M}}_l^{(2)}(\mathbf{b})
,
\label{sec3a5}
\end{eqnarray}
where
\begin{equation}
{\mathbf{M}}_k^{(1)}(\mathbf{a})={\mathbf{M}}_k(\mathbf{a})\otimes\openone
\quad\hbox{and}\quad
{\mathbf{M}}_l^{(2)}(\mathbf{b})=\openone\otimes{\mathbf{M}}_l(\mathbf{b})
.
\label{sec3a6}
\end{equation}
In Sec.~\ref{sec2j}, we gave a proof that
the matrix $\mathbf{F}(P,N)$ describing the source of the SG experiment is positive semidefinite.
Using the same reasoning, it follows that $\mathbf{F}(P,N)$ appearing in Eq.~(\ref{sec3a5})
is positive semidefinite as well.

With the help of the basis in Eq.~(\ref{sec2j0}), we can write down the most general expression of $\mathbf{F}(P,N)$ as
\begin{eqnarray}
\mathbf{F}(P,N)&=&\sum_{i,j=0}^8 f_{i,j}\mathbf{B}_i\otimes\mathbf{B}_j
,
\label{sec3a7}
\end{eqnarray}
where the expansion coefficients $f_{i,j}=f^\ast_{j,i}$ can, in principle, be determined from the data Eq.~(\ref{sec3a5}),
obtained by making experiments with several different choices of $(\mathbf{a},\mathbf{b})$.
However, in practice, the experimental procedure to determine the 80 real numbers entering Eq.~(\ref{sec3a7}) is quite cumbersome.
In contrast, given a specific expression for $\mathbf{F}(P,N)$, it is straightforward to compute
the moments $\langle k^p l^q\rangle$.
For instance, if we choose
\begin{eqnarray}
\mathbf{F}(P,N)=
\frac{1}{3}
\left(\begin{array}{rrrrrrrrr}
\phantom{-}0 & \phantom{-}0 & \phantom{-}0 & \phantom{-}0 & \phantom{-}0 & \phantom{-}0 & \phantom{-}0  & \phantom{-}0 & \phantom{-}0\\
\phantom{-}0 & \phantom{-}0 & \phantom{-}0 & \phantom{-}0 & \phantom{-}0 & \phantom{-}0 & \phantom{-}0  & \phantom{-}0 & \phantom{-}0\\
\phantom{-}0 & \phantom{-}0 & \phantom{-}1 & \phantom{-}0 & -1 & \phantom{-}0 & \phantom{-}1& \phantom{-}0 & \phantom{-}0\\
\phantom{-}0 & \phantom{-}0 & \phantom{-}0 & \phantom{-}0 & \phantom{-}0 & \phantom{-}0 & \phantom{-}0& \phantom{-}0 & \phantom{-}0\\
\phantom{-}0 & \phantom{-}0 & -1 & \phantom{-}0 & \phantom{-}1 & \phantom{-}0 & -1& \phantom{-}0 & \phantom{-}0\\
\phantom{-}0 & \phantom{-}0 & \phantom{-}0 & \phantom{-}0 & \phantom{-}0 & \phantom{-}0 & \phantom{-}0& \phantom{-}0 & \phantom{-}0\\
\phantom{-}0 & \phantom{-}0 & \phantom{-}1 & \phantom{-}0 & -1 & \phantom{-}0 & \phantom{-}1& \phantom{-}0 & \phantom{-}0\\
\phantom{-}0 & \phantom{-}0 & \phantom{-}0 & \phantom{-}0 & \phantom{-}0 & \phantom{-}0 & \phantom{-}0  & \phantom{-}0 & \phantom{-}0\\
\phantom{-}0 & \phantom{-}0 & \phantom{-}0 & \phantom{-}0 & \phantom{-}0 & \phantom{-}0 & \phantom{-}0  & \phantom{-}0 & \phantom{-}0\\
\end{array}\right)
,
\label{sec3a8}
\end{eqnarray}
which, in quantum theory language, represents the pure state
$|\Psi\rangle=\left(|-1,1\rangle - |0,0\rangle + |+1,-1\rangle \rangle\right)/\sqrt{3}$
with total spin zero, we obtain
\begin{eqnarray}
\langle k \rangle &=& \mathbf{Tr\;}\mathbf{F}(P,N) \;(\mathbf{a}\cdot\mathbf{S}_1)=\langle \mathbf{a}\cdot\mathbf{S_1}\rangle=0,
\nonumber \\
\langle l \rangle &=&\mathbf{Tr\;}\mathbf{F}(P,N) \;(\mathbf{b}\cdot\mathbf{S}_2)=\langle \mathbf{b}\cdot\mathbf{S_2}\rangle=0,
\nonumber \\
\langle k l \rangle &=& \mathbf{Tr\;}\mathbf{F}(P,N)\; (\mathbf{a}\cdot\mathbf{S}_1) \;(\mathbf{b}\cdot\mathbf{S}_2)
= \langle \mathbf{a}\cdot\mathbf{S_1}\;\mathbf{b}\cdot\mathbf{S_2}\rangle
= -\frac{2}{3}\; \mathbf{a}\cdot\mathbf{b},
\nonumber \\
\langle k^2 l \rangle &=&
\mathbf{Tr\;}\mathbf{F}(P,N) \;(\mathbf{a}\cdot\mathbf{S_1})^2\;(\mathbf{b}\cdot\mathbf{S_2})=
\langle (\mathbf{a}\cdot\mathbf{S_1})^2\;(\mathbf{b}\cdot\mathbf{S_2})\rangle=0,
\nonumber \\
\langle k l^2 \rangle &=&
\mathbf{Tr\;}\mathbf{F}(P,N) \;(\mathbf{a}\cdot\mathbf{S_1})\;(\mathbf{b}\cdot\mathbf{S_2})^2=
\langle (\mathbf{a}\cdot\mathbf{S_1})\;(\mathbf{b}\cdot\mathbf{S_2})^2\rangle=0,
\nonumber \\
\langle k^2 l^2 \rangle &=&\mathbf{Tr\;}\mathbf{F}(P,N) \;(\mathbf{a}\cdot\mathbf{S_1})^2\;(\mathbf{b}\cdot\mathbf{S_2})^2
=\langle (\mathbf{a}\cdot\mathbf{S_1})^2\;(\mathbf{b}\cdot\mathbf{S_2})^2\rangle
=\frac{1}{3} \left(1+ (\mathbf{a}\cdot\mathbf{b})^2\right)
.
\label{sec3a9}
\end{eqnarray}
From Eq.~(\ref{sec3a9}) it is clear that all the moments are invariant for arbitrary rotations
of the laboratory reference frame, i.e. the matrix Eq.~(\ref{sec3a8}) describes
a source which emits particles with properties that, upon measurement, do not change if we rotate the source.

\section{Discussion}\label{sec11}

Quantum theory allows us to describe situations in which we are unable to predict each individual event
but are able to represent the collection of such events by relative frequencies only.
In contrast, the key of our construction of the quantum formalism is to start from the notion of individual events.
Therefore, by construction, this formulation of quantum theory is free of the usual interpretational issues
related to the meaning of the density matrix/wave function and other mathematical tools, such as probability theory.
In our treatment, the events are the ``real thing'' and the wave function only serves
as a mathematical vehicle to represent the observed frequencies.

Our construction of the mathematical framework that forms the basis of quantum theory
starts with the consistent application of the general and natural idea of separating descriptions
in parts and of a simple rewriting of the representation of the frequencies of observed events.
For concreteness, we presented an explicit construction of the quantum theoretical description for the case of
the SG and EPRB experiment with three possible outcomes per particle.
There is nothing in our explicit treatments that prevents generalizations
to an arbitrary number of outcomes per particle and an arbitrary number of particles.

Therefore, it may be useful to discuss the relation between the various steps in
our explicit construction and the commonly accepted axiomatic formulation of quantum theory.
The latter is well documented~\cite{NEUM55,BALL03,WEIN03,KHRE09}.
For the purpose of discussing the relation with the construction given
in this paper, the formulation given in Ref.~\cite{BALL03} is most convenient.
We list each of them together with a reference to the point in this paper where they appear.
\begin{enumerate}
\item
{\sl To each dynamical variable R (physical concept) there corresponds
a linear operator $R$ (mathematical object), and the possible values of the
dynamical variable are the eigenvalues of the operator~\cite{BALL03}.}

Physical concepts ultimately relate to sense impressions.
As a metaphor for these impressions, we use the different outcomes of a (double) SG or EPRB experiment.
The elementary mathematical objects are the projection operators $\mathbf{M}_k(\mathbf{e}_z)$, see Sec.~\ref{sec2g}.
More complicated mathematical objects can be constructed by appropriate linear combinations of these
projection operators, exactly as in quantum theory.
\medskip
\item
{\sl To each state there corresponds a unique state operator. The
average value of a dynamical variable R, represented by the operator $R$, in
the virtual ensemble of events that may result from a preparation procedure
for the state, represented by the operator $\rho$, is
$\langle R\rangle = \mathbf{Tr\;}\rho R /\mathbf{Tr\;}\rho$~\cite{BALL03}.}

The unique state operator, denoted by the matrix $\mathbf{F}(P,N)$ appears after
separating the description of the data set into a description of the particle(s) and SG magnet(s).
In Sec.~\ref{sec2j}, we show that the usual properties ($\mathbf{Tr\;}\mathbf{F}(P,N)=1$
and $\mathbf{F}(P,N)$ non-negative definite) follow
from the fact that the numbers of events are non-negative numbers.
The expression of the average value of a dynamical variable R follows
directly from the requirement that the description separates, see Sec.~\ref{sec2e}.
There is no need to consider a virtual ensemble of events.
\end{enumerate}

Our derivation of the von Neumann equation and Schr\"odinger equation, see Sec.~\ref{sec4a},
builds on the theorem that the trace of a matrix is zero if and only if the matrix can be
written as a commutator~\cite{SHOD36,ALBE57}
and the condition that
the eigenvalues of $\mathbf{F}(P,N,\lambda)$ are independent of $\lambda$.
\POS\ applied to the relative frequencies of events and elementary use of matrix algebra, together
with standard assumptions about the space-time continuum,
are sufficient to construct the mathematical framework of quantum theory.
But even with these additional assumptions, we have $\mathrm{SOC}\models \mathrm{QT} $.

\section{Conclusion}\label{sec10}

We have explored a route to construct the mathematical framework of quantum theory
without relying on the accepted set of quantum physics postulates.
The Stern-Gerlach and EPRB experiment, both key to the development of quantum theory,
have been used to demonstrate
that the basic postulates of the quantum formalism follow from describing
 the number of particles in the outgoing beams
in terms of separate descriptions of the individual components that make up the experiment.
The \POS\ approach readily handles any value of the number of outcomes.
The Schr\"odinger and the von Neumann equation, two equations governing the time evolution of quantum systems,
are shown to model time-dependent data, the description of which can be separated in parts.

The general message of this paper may be summarized as follows.
The idea that the description of a quantum physics experiment can be decomposed
in descriptions of independent parts (e.g. preparation and measurement stage) is
not only an implicit assumption in standard formulations of quantum theory but is, as we show in this paper,
already sufficient to expose its basic mathematical structure embodied in Eq.~(\ref{sec2e3a}) and generalizations thereof.

However, \POS\ itself does not suffice to derive, for each individual experiment, the concrete, explicit descriptions
that we know from quantum theory.
To this end, \POS\ has to be supplemented with standard assumptions about the symmetries of the space-time continuum
and, as in the case of the time-dependent Schr\"odinger equation, with other assumptions as well.
In other words, \POS\ can be used to describe experiments performed under different but separable conditions
which may or may not be describable by the quantum formalism.
In any case, the \POS-based construction of the quantum formalism explains the success of quantum theory as a tool
to describe the statistics of a vast amount of (quantum or non-quantum)
experiments for which we have no means to predict individual events.

Finally, we believe that the approach of introducing the quantum theoretical framework pursued in this paper
may contribute to its demystification
because there (i) is no need to motivate the postulates P1 and P2.
and (ii) it is void of the usual postulates/interpretations regarding ``wave functions'', ``observables'', ``quantization rules'',
``Born's rule'', ``probabilities'', and the like.

\section*{Acknowledgements}
The work of M.I.K. is supported by the European Research Council (ERC) Advanced Grant No. 338957 FEMTO/NANO.
D.W. is supported by the Initiative and Networking Fund of the Helmholtz Association through the Strategic Future
Field of Research project ``Scalable solid state quantum computing (ZT-0013)''.

\appendix
\section{A. Rotation of the SG magnet}{\label{APP2}}

According to Rodrigues' formula, rotating a unit vector $\mathbf{u}$ about
the axis of rotation $\bm\alpha$ (with $\Vert \bm\alpha \Vert=1$) by an angle $\phi$ yields the vector
\begin{eqnarray}
\mathbf{v}=\mathbf{u}\cos\phi+(\bm\alpha\times\mathbf{u})\sin\phi + (\bm\alpha\cdot\mathbf{u})\bm\alpha (1-\cos\phi)
=\mathbf{u}+(\bm\alpha\times\mathbf{u})\sin\phi + \bm\alpha\times(\bm\alpha\times\mathbf{u})(1-\cos\phi)
.
\label{sec2c2}
\end{eqnarray}
Conversely, if $\mathbf{u}$ and $\mathbf{w}$ are unit vectors, setting
\begin{eqnarray}
\bm\alpha=\frac{\mathbf{u}\times\mathbf{w}}{\Vert \mathbf{u}\times\mathbf{w} \Vert}
=\frac{1}{\sin\phi} \mathbf{u}\times\mathbf{w}
\quad,\quad \cos\phi=\mathbf{u}\cdot\mathbf{w}
,
\label{sec2c2a}
\end{eqnarray}
defines the rotation about the unit vector $\bm\alpha$ by the angle $\phi$
which changes $\mathbf{u}$ into $\mathbf{w}$.

We now ask what happens to the projection of the spin matrices $\mathbf{S}$ on the unit vector $\mathbf{u}$
if we perform the same rotation in spin-space as the one that changes $\mathbf{u}$ into $\mathbf{w}$.
To answer this question, we introduce the operator
\begin{eqnarray}
g(\phi)\equiv e^{-i\,\phi \bm\alpha\cdot\mathbf{S}}\; \mathbf{u}\cdot\mathbf{S}\; e^{+i\,\phi\bm\alpha\cdot\mathbf{S}}
,
\label{sec2c3}
\end{eqnarray}
and using the commutation relations of the angular momentum (spin) operators $[S^x,S^y]=iS^z$,
$[S^z,S^x]=iS^y$, and $[S^y,S^z]=iS^z$ we find
\begin{eqnarray}
\frac{\partial g(\phi)}{\partial \phi}&=&
e^{-i\,\phi \bm\alpha\cdot\mathbf{S}}\; (\bm\alpha\times\mathbf{u})\cdot\mathbf{S}\; e^{+i\,\phi\bm\alpha\cdot\mathbf{S}}
,
\\
\frac{\partial^2 g(\phi)}{\partial \phi^2}&=&
e^{-i\,\phi \bm\alpha\cdot\mathbf{S}}\; \bm\alpha\times(\bm\alpha\times\mathbf{u})\cdot\mathbf{S}\; e^{+i\,\phi\bm\alpha\cdot\mathbf{S}}
=e^{-i\,\phi \bm\alpha\cdot\mathbf{S}}\; (\bm\alpha\cdot\mathbf{u})\bm\alpha\cdot\mathbf{S}\; e^{+i\,\phi\bm\alpha\cdot\mathbf{S}}
-e^{-i\,\phi \bm\alpha\cdot\mathbf{S}}\; \mathbf{u}\cdot\mathbf{S}\; e^{+i\,\phi\bm\alpha\cdot\mathbf{S}}
\nonumber\\
&=&-g(\phi)
+(\bm\alpha\cdot\mathbf{u}) e^{-i\,\phi \bm\alpha\cdot\mathbf{S}}\; \bm\alpha\cdot\mathbf{S}\; e^{+i\,\phi\bm\alpha\cdot\mathbf{S}}
=-g(\phi) +(\bm\alpha\cdot\mathbf{u})\bm\alpha\cdot\mathbf{S}
.
\label{sec2c4}
\end{eqnarray}
Integrating the second-order differential equation Eq.~(\ref{sec2c4}) yields
\begin{eqnarray}
g(\phi)&=&g(0)\cos\phi+g'(0)\sin\phi + (\bm\alpha\cdot\mathbf{u})\bm\alpha\cdot\mathbf{S} (1-\cos\phi)
\nonumber\\
&=&
\mathbf{u}\cdot\mathbf{S}\cos\phi
+(\bm\alpha\times\mathbf{u})\cdot\mathbf{S}\sin\phi + (\bm\alpha\cdot\mathbf{u})\bm\alpha\cdot\mathbf{S} (1-\cos\phi)
\nonumber\\
&=&
\left[\mathbf{u}\cos\phi
+(\bm\alpha\times\mathbf{u})\sin\phi + (\bm\alpha\cdot\mathbf{u})\bm\alpha (1-\cos\phi)
\right]\cdot\mathbf{S}=\mathbf{w}\cdot\mathbf{S}
.
\label{sec2c5}
\end{eqnarray}
In other words, we have shown that a rotation in 3D space that changes
a unit vector $\mathbf{u}$ into a unit vector $\mathbf{w}$
corresponds to a rotation in spin-space that changes
the projection of the spin on the direction $\mathbf{u}$
to
the projection of the spin on the direction $\mathbf{w}$
according to
\begin{eqnarray}
\mathbf{w}\cdot\mathbf{S}=e^{-i\,\phi \bm\alpha\cdot\mathbf{S}}\; \mathbf{u}\cdot\mathbf{S}\; e^{+i\,\phi\bm\alpha\cdot\mathbf{S}}
.
\label{sec2c6}
\end{eqnarray}
\section{B. Projection operators and moment expansions}{\label{APP3}}
We give a general proof that explicit expressions
for the projectors onto the eigenspaces of a non-degenerate Hermitian matrix $A$
can be obtained by expanding an arbitrary function of the eigenvalues of $A$ in
terms of its moments, and then symbolically replacing the $p$th moment by the pth
power of $A$.

Let $A$ be a $N\times N$ Hermitian matrix with non-degenerate eigenvalues $\lambda_1,\ldots,\lambda_N$
and corresponding eigenvectors $v_1,\ldots,v_N$.
The matrix
\begin{eqnarray}
P_i(A)&=&\prod_{j\not= i} \frac{A-\openone \lambda_j}{\lambda_i-\lambda_j} = P_i^\dagger(A)
,
\label{app3a}
\end{eqnarray}
satisfies $P_i(A) v_k=\delta_{i,k} v_k$ and $P_j(A)P_i(A) v_k=\delta_{i,k} \delta_{j,k} v_k$
and is therefore a projector on the one-dimensional space defined by the eigenvector $v_k$.
Formally expanding the product in Eq.~(\ref{app3a}), we obtain
\begin{eqnarray}
P_i(A)&=&\sum_{n=1}^N b_{i,n} A^{n-1}
.
\label{app3b}
\end{eqnarray}
From the same formal expansion of the real-valued function defined by
\begin{eqnarray}
g_i(\lambda)&=&\prod_{j\not= i} \frac{\lambda-\openone \lambda_j}{\lambda_i-\lambda_j}=\sum_{n=1}^N b_{i,n} \lambda^{n-1}
,
\label{app3c}
\end{eqnarray}
it follows that
\begin{eqnarray}
g_i(\lambda_k)&=&\delta_{i,k}=\sum_{n=1}^N b_{i,n} \lambda_k^{n-1}
.
\label{app3d}
\end{eqnarray}
Introducing the Vandermonde matrix $V_{k,n}=\lambda_k^{n-1}$, Eq.~(\ref{app3d}) reads $b V^\mathrm{T}=\openone$
or, equivalently, $b=\left(V^\mathrm{T}\right)^{-1}$ where
the assumption that the eigenvalues are non-degenerate guarantees that the inverse of $V^\mathrm{T}$ exists.
Therefore we may write Eq.~(\ref{app3b}) as
\begin{eqnarray}
P_i(A)&=&\sum_{n=1}^N \left(V^\mathrm{T}\right)^{-1}_{i,n} A^{n-1}
.
\label{app3e}
\end{eqnarray}

On the other hand, the moments $m_p$ of the function $f(\lambda_1),\ldots,f(\lambda_N)$ are defined as
\begin{eqnarray}
m_p&=&\sum_{j=1}^N  f(\lambda_j) \lambda_j^{p-1} = \sum_{j=1}^N  f(\lambda_j) V_{j,p}
\nonumber\\
\noalign{\noindent or}
M &=& V^\mathrm{T} F
,
\label{app3f}
\end{eqnarray}
where $M=(m_1,\ldots,m_{N})^\mathrm{T}$ and $F=(f(\lambda_1),\ldots,f(\lambda_N))^\mathrm{T}$.
From Eq.~(\ref{app3f}) it follows directly that
\begin{eqnarray}
F &=& \left(V^\mathrm{T}\right)^{-1} M
\nonumber\\
\noalign{\noindent or}
f(\lambda_i)&=&\sum_{p=1}^N \left(V^\mathrm{T}\right)^{-1}_{i,p} m_{p-1}
.
\label{app3g}
\end{eqnarray}
Replacing the symbol $m_{p-1}$ in Eq.~(\ref{app3g}) by $A^{p-1}$,
the right-hand-side of Eq.~(\ref{app3g}) becomes identical to the
right-hand-side of Eq.~(\ref{app3e}), which proves the statement made in the beginning
of this section.

\bibliography{../../all18}
\end{document}